\documentclass[amsmath,amssymb,aps,eqsecnum]{revtex4}
\usepackage{amsfonts,amsthm}
\usepackage{epsfig}
\usepackage{color}
\newcommand{\Tr}{{\rm{Tr}}}
\newcommand{\SpN}{{\rm{Sp}}_{2N}(\mathbb{R})}
\newcommand{\spN}{{\mathfrak{sp}}_{2N}(\mathbb{R})}
\begin{document}
\title{Energy correlations for a random matrix model of disordered bosons}

\author{T.\ Lueck$^1$, H.-J.\ Sommers$^2$, and M.R.\ Zirnbauer$^1$}
\affiliation {$^1$ Institut f\"{u}r Theoretische Physik,
Universit\"{a}t zu K\"{o}ln, Z\"{u}lpicher Stra{\ss}e 77, 50937
K\"{o}ln, Germany,}
\affiliation{$^2$Fachbereich Physik, Universit\"{a}t Duisburg-Essen,
Campus Essen, 45117 Essen, Germany}
\date{July 10, 2006}
\begin{abstract}
Linearizing the Heisenberg equations of motion around the ground
state of an interacting quantum many-body system, one gets a
time-evolution generator in the positive cone of a real symplectic
Lie algebra. The presence of disorder in the physical system
determines a probability measure with support on this cone. The
present paper analyzes a discrete family of such measures of
exponential type, and does so in an attempt to capture, by a simple
random matrix model, some generic statistical features of the
characteristic frequencies of disordered bosonic quasi-particle
systems. The level correlation functions of the said measures are
shown to be those of a determinantal process, and the kernel of the
process is expressed as a sum of bi-orthogonal polynomials. While the
correlations in the bulk scaling limit are in accord with sine-kernel
or GUE universality, at the low-frequency end of the spectrum an
unusual type of scaling behavior is found.
\end{abstract}
\maketitle

\section{Introduction}\label{sect:I}

Perturbing the ground state of an interacting quantum many-body
system and linearizing the Heisenberg equations of motion for the
boson Fock operators, one faces the standard problem of small
oscillations. Concrete examples are furnished by the vibrational
modes of a solid, the spin waves in a magnet, the electromagnetic
modes in an optical medium, and the oscillations of the superfluid
density of a Bose-Einstein condensate. Common to these excitations is
that they second-quantize as bosons or bosonic quasi-particles.

Adding some amount of disorder to the system, one may ask: what are
the statistical features of the excitation spectrum and, in
particular, which of these features (if any) reflect the bosonic
nature of the quasi-particle excitations? Is there some kind of
universality akin to the Wigner-Dyson universality known from other
disordered systems? If so, what are the universal laws, and what's
the role of symmetry in determining these laws?

In the parallel case of fermionic quasi-particles the situation is
now fairly well understood. If the system is of metallic type and in
the ergodic limit, the statistical behavior at high energies is in
accord with the universal laws of Wigner-Dyson statistics. For low
excitation energies, however, the canonical anti-commutation
relations obeyed by the fermion operators make themselves felt: they
constrain the form of the Hamiltonian matrix and thus give rise to
several new universality classes beyond Dyson's threefold way
\cite{dyson}.  Some of these are realized by chiral Dirac fermions in
a random gauge field \cite{jjmv}, others by quasi-particles in
disordered gapless superconductors \cite{az,asz}. A complete symmetry
classification of quadratic fermion Hamiltonians has been carried out
\cite{hhz}, and the role of Riemannian symmetric spaces and
superspaces in providing an effective description has been emphasized
\cite{suprev,cm}.

Progress has been slower for bosonic systems, and so for good reason,
as these are set apart by several distinctive features from other
random problems studied and solved in the past. For one thing, in the
case of bosons it makes little sense to choose -- as one often does
for fermions -- the matrix elements of the quasi-particle Hamiltonian
as independent and identically distributed random variables. In fact,
most of the boson Hamiltonians produced in such a manner would
generate runaway dynamics rather than oscillatory motion around a
stable ground state. In the case of bosons one therefore has to pay
attention to the fact that the matrix elements depend in a
complicated way on the ground state of the many-boson system and
hence on the disorder of the microscopic parent problem
\cite{cg1,cg2}. As a technical consequence, a direct analog of the
so-called Gaussian Ensembles, which were pivotal in initiating the
Wigner-Dyson theory and establishing its universal statistics, is
unavailable in the context of bosons.

For another complication, low-frequency bosons are usually
insensitive to weak disorder. Many of the excitations listed above
are Goldstone bosons associated with a spontaneously broken symmetry,
and for such excitations low frequency is tantamount to low wave
number, or large wave length, which causes the scattering by disorder
to be suppressed, as the disorder is effectively seen only on average
over regions of size given by the large wave length. Thus the
disorder averages out and becomes less effective, and hence the
behavior of weakly disordered Goldstone bosons tends to be
system-specific. (Of course this still leaves it possible for weakly
disordered bosons of \emph{non}-Goldstone type to exhibit universal
statistics \cite{cg1,cg2}.) In order for any universality to set in,
the disorder strength often has to be so large that standard
calculational tools such as the impurity diagram technique fail to
apply.

In the present paper we are going to introduce and completely solve a
simple random matrix model of disordered bosonic quasi-particles,
which we believe to be most closely analogous to the Wigner-Dyson
Gaussian Ensembles while retaining the crucial features of bosonic
statistics and stability of the motion. In a follow-up paper we will
investigate the question whether this simple model might be
representative of a whole universality class of related problems.

To formulate the model, let $q_j\, , p_j$ $(j = 1, \ldots, N)$ be a
canonical set of position and momentum operators, and consider their
linearized Heisenberg equations of motion in the most general form:
\begin{displaymath}
    \dot{q}_j = \sum_{i=1}^N (q_i A_{ij} + p_i C_{ij})\;,\qquad
    - \dot{p}_j = \sum_{i=1}^N (q_i B_{ij} + p_i A_{ji})\;,
\end{displaymath}
where $B_{ij} = B_{ji}\,$, $C_{ij} = C_{ji}\,$, and $A_{ij}$ are real
numbers. If the system was invariant under time reversal ($q_j
\mapsto q_j\,$, $p_j \mapsto -p_j$), the coefficients $A_{ij}$ would
have to be zero, but we here consider the generic case without
symmetries. The criterion for stability of the dynamics is that the
stability matrix be positive:
\begin{displaymath}
    h := \begin{pmatrix} B & A \\ A^t & C \end{pmatrix} > 0 \;.
\end{displaymath}
Assuming $h^t = h > 0\,$, the generator of the Heisenberg time
evolution,
\begin{displaymath}
    X := \begin{pmatrix} A &-B \\ C &-A^t \end{pmatrix} \;,
\end{displaymath}
has eigenvalues that come as imaginary pairs $\pm \mathrm{i}
\omega_j$ where $\omega_j > 0$ ($j = 1, \ldots, N$) are the
characteristic frequencies (or single-boson energies) of the
small-amplitude motion. In a classical setting one would introduce
the generator $X$ as the symplectic gradient of the Hamiltonian
function linearized at a stable equilibrium point of the classical
flow.

The natural transformation group of the problem at hand is the real
symplectic group in $2N$ dimensions, $\mathrm{Sp}_{2N}(\mathbb{R})$,
acting by linear canonical transformations on the operators $q_j\, ,
p_j$ and by conjugation on the generator $X\,$. We can now explain
one of the distinctive features of the present problem: when
formulating the Gaussian Ensembles of the Wigner-Dyson theory one
makes the postulate that the transformation group of the problem
($\mathrm{O}_N$, $\mathrm{U}_N$, or $\mathrm{USp}_{2N}$, as the case
may be) also be the symmetry group of the chosen probability measure,
whereas in our case no such simplification is possible. Indeed,
$\mathrm{Sp}_{2N}(\mathbb{R})$ is non-compact, and a probability
measure $d\mu$ cannot be invariant under a non-compact group action
and at the same time have total mass $\int d\mu = 1$.

One is therefore looking for some construction principle other than
symmetry. Our key here is the positivity of the real symmetric
stability matrix $h:$ a natural way of building positive real
symmetric matrices $h$ is by adding a sufficient number of rank-one
projectors with positive weights. Equivalently, we may put
\begin{equation}\label{eq:1.1}
    h_{ij} = \sum_{\alpha = 1}^M v_{i \alpha} v_{j \alpha} \qquad
    (i,j = 1, \ldots, 2N)
\end{equation}
for some set of real numbers $v_{i \alpha}\,$. We now consider the
$v_{i \alpha}$ as the fundamental variables, and choose them to be
independent and normal (or Gaussian) distributed random variables
with zero mean and variance $\tau^{-1}$. Then we use (\ref{eq:1.1})
to push forward the normal distribution for the $v_{i \alpha}$ to a
probability distribution $d\mu(h)$ for $h$ (and hence for $X$). If $M
\ge 2N$, the result is
\begin{equation}\label{eq:1.2}
    d\mu(h) \propto \mathrm{e}^{-\frac{1}{2} \tau \mathrm{Tr}\, h}\,
    \mathrm{Det}(h)^{\frac{1}{2}(l-1)} \, \prod\nolimits_{i \le j}
    dh_{ij} \;, \qquad l = M - 2 N \ge 0 \;,
\end{equation}
with the domain for $h$ still defined by $h > 0\,$. The probability
distribution (\ref{eq:1.2}) is the object of study of this paper.

We now give a summary of the contents and the results of the paper.
After collecting some basic facts from symplectic linear algebra in
Sect.\ \ref{sect:II}, we reduce $d\mu(h)$ in Sect.\ \ref{sect:III} to
a probability distribution on the space of characteristic frequencies
$\omega_1 , \ldots, \omega_N$ (the positive eigenvalues of
$-\mathrm{i}X$), and find this to be
\begin{equation}
    d\mu_{N,l}(\omega_1, \ldots, \omega_N) = c_{N,l}(\tau)
    \prod\limits_{i < j} (\omega_i - \omega_j) (\omega_i^2 - \omega_j^2)
    \prod\limits_{k =1}^N \omega_k^l \, \mathrm{e}^{-\tau \omega_k}\,
    d\omega_k \;.
\end{equation}
Using the method of bi-orthogonal polynomials we show in Sect.\
\ref{sect:V.C} that the $n$-level correlation functions of this
probability distribution are of determinant type and are completely
determined in the usual way -- see (\ref{eq:correls}) -- by a certain
kernel $K_N(\omega, \tilde\omega)$ given as a sum over bi-orthogonal
polynomials. We compute the large-$N$ asymptotics of this kernel in
the bulk of the spectrum (in Sect.\ \ref{sect:V.D}) and at the `hard'
edge $\omega = 0$ (Sect.\ \ref{sect:V.E}), using a contour integral
representation of the bi-orthogonal polynomials (Sect.\
\ref{sect:V.C}). In the former case we establish the scaling limit
\begin{equation}\label{eq:1.4}
    \tau \lim_{N \to \infty} K_N(Nx/\tau + \omega\, , Nx/\tau
    + \tilde\omega) = \frac{\sin\big(\pi \rho_\infty(x) (\omega -
    \tilde\omega)\big)}{\pi(\omega - \tilde\omega)}\, \mathrm{e}^{
    -r(x)(\omega - \tilde\omega)} \;,
\end{equation}
which is independent of $l\,$. The function $\rho_\infty(x)$ of the
scaling variable $x = \omega\tau / N$ is the large-$N$ limit of the
level density. Viewing $\pi \rho_\infty(x)$ as the imaginary part of
a Green's function $\lim_{\epsilon \to 0+} g(x+\mathrm{i}\epsilon)$,
the function $r(x)$ is the real part. We compute $\rho_\infty(x)$ by
two independent methods (from a variational calculation in Sect.\
\ref{sect:cubic}, and from bi-orthogonal polynomials in Sect.\
\ref{sect:V.D}), with the result being
\begin{equation}\label{eq:1.5}
    \rho_\infty(x) = \frac{\tau}{2\pi} (x/b)^{-1/3} \big(
    (1 + \sqrt{1-x^2/ b^2})^{1/3} - (1 - \sqrt{1-x^2/b^2})^{1/3}
    \big) \qquad (0 < x \le b = 3 \sqrt{3})\;.
\end{equation}
Apart from the last factor, which is irrelevant since it cancels on
passing to the level correlation functions, the right-hand side of
(\ref{eq:1.4}) is the famous sine kernel known from systems with
unitary symmetry. Thus we recover Wigner-Dyson universality of the
class of the Gaussian Unitary Ensemble (GUE) at bulk frequencies.

At low frequencies $\omega \sim N^{-1/2}$ we find convergence to an
unusual kind of scaling limit:
\begin{equation}\label{eq:1.6}
    \lim_{N\to\infty} N^{-1/2} K_N(N^{-1/2} y_1 / \tau, N^{-1/2}
    y_2 / \tau ) = \frac{\tau^2}{2\pi^2} \int\limits_{\mathrm{i}
    \mathbb{R} + \epsilon} du \oint\limits_{\mathrm{U}_1}
    \frac{dv}{v} \,\, \mathrm{e}^{- y_1 / u + y_2 / v}\, (v/u)^l \,
    \frac{\mathrm{e}^{u^2 - v^2} - 1}{u^2 - v^2} \;,
\end{equation}
where $\mathrm{U}_1$ denotes the unit circle in $\mathbb{C}\,$, and
$\mathrm{i}\mathbb{R} + \epsilon$ is any axis in the right half plane
parallel to the imaginary axis. The result (\ref{eq:1.6}) is
reminiscent of formulas obtained by Efetov's supersymmetry method,
with $u$ and $v$ playing the role of radial polar coordinates of a
Riemannian symmetric superspace. We intend to elucidate this
connection in a future publication.

\section{The Hamiltonians of stable motions}
\label{sect:II}

Let there be some position variables $q_1, \ldots, q_N$ and canonical
momenta $p_1, \ldots, p_N$\,, and consider Hamiltonians $H$ of the
quadratic form
\begin{equation}\label{Hamiltonian}
    H = \frac{1}{2}\sum_{i,j=1}^{N} \big( C_{ij}\, p_i p_j + B_{ij}\,
    q_i q_j + A_{ij} (q_i p_j + p_j q_i) \big) \,,
\end{equation}
where $A$, $B$, and $C$ are real matrices satisfying $B = B^t$ and $C
= C^t$. Rewriting $H$ as
\begin{displaymath}
    H = \frac{1}{2}(\begin{array}{cc} q & p \end{array})
    \left(\begin{array}{cc} A & -B \\ C & - A^t\end{array}\right)
    \left(\begin{array}{c} p \\ -q \end{array}\right) \;,
\end{displaymath}
we see that the matrix, $X$, of $H$ satisfies the linear condition
\begin{equation}\label{eq:sympl-Lie}
    X^t J + J X = 0 \;, \qquad J = \begin{pmatrix} 0 &-1_N \\
    1_N& 0\end{pmatrix} \;, \quad X = \begin{pmatrix} A &-B\\ C
    &-A^t\end{pmatrix} \;.
\end{equation}
This is saying that $X$ lies in $\mathfrak{sp}_{2N}(\mathbb{R})$, the
Lie algebra of the real symplectic group defined by
\begin{displaymath}
    \SpN = \{g\in {\rm GL}_{2N}({\mathbb{R}})\, |\, g^t J g = J\} \;.
\end{displaymath}
A matrix $X \in \spN$ need not be diagonalizable (e.g., the generator
of free motion, $A = B = 0$ and $C = 1_N\,$, isn't); and even if it
is, the eigenvalues will in general be complex.

We now impose the condition
\begin{equation}
    h := \begin{pmatrix} B & A \\ A^t & C \end{pmatrix} > 0 \;,
\end{equation}
i.e., we require all eigenvalues of the real symmetric matrix $h$ to
be positive. The corresponding domain in $\spN$ will be denoted by
$\mathfrak{E}^0$:
\begin{equation}
    \mathfrak{E}^0 := \{ X \in \spN \, | \, X = h J \, ,
    \, h = h^t > 0 \} \;.
\end{equation}
Although the eigenvalues of $h$ have no direct relation to the
dynamics of the system, positivity of $h$ ensures that the motion
generated by the Hamiltonian $H$ is stable, or `elliptic'. As a
consequence of ellipticity, there exists some linear canonical
transformation $(\mathbf{q}\, , \mathbf{p}) \to (\mathbf{Q}\,
,\mathbf{P})$ which takes the Hamiltonian to a sum of harmonic
oscillators,
\begin{displaymath}
    H = \frac{1}{2}\sum_{i=1}^{N} \big( P_i^2 + \omega_i^2 Q_i^2 )\;,
\end{displaymath}
with $\omega_i^2 > 0\,$. Put differently, for $X \in \mathfrak{E}^0$
one can always find a symplectic transformation $g \in \SpN$ that
conjugates $X$ to quasi-diagonal form:
\begin{equation}\label{eq:X-diag}
    X = g \Omega g^{-1}\;, \quad \Omega = \begin{pmatrix} 0 &-\omega
    \\ \omega &0 \end{pmatrix}\;, \quad \omega = \mathrm{diag}
    (\omega_1 , \omega_2 , \ldots, \omega_N) \;,
\end{equation}
with real and positive $\omega_i$ $(i = 1, \ldots, N)$.

All of the discussion below will be based on the elliptic domain
$\mathfrak{E}^0$.  Let us therefore collect some of its mathematical
properties. First of all, if $X$ is in $\mathfrak{E}^0$, then so is
its conjugate $g X g^{-1}$ by any element $g \in \SpN$. Thus
$\mathfrak{E}^0$ is invariant under the action of $\SpN$ on
$\mathfrak{E}^0$ by conjugation. Secondly, let $\mathfrak{t}$ denote
the Abelian algebra of block-diagonal matrices of the form of
$\Omega$ in (\ref{eq:X-diag}) but with diagonal elements $\omega_i$
that are any real numbers (not necessarily positive). Let
$\mathfrak{t}_+ \subset \mathfrak{t}$ be the subset of block-diagonal
$\Omega$ with positive $\omega_i\,$. Then, as we said earlier, every
$X \in \mathfrak{E}^0$ is conjugate to a unique $\Omega \in
\mathfrak{t}_+$ by some $g \in \SpN$.  Thirdly, introducing $T :=
\exp( \mathfrak{t})$, which is an $N$-dimensional compact torus, $T
\cong (\mathrm{S}^1)^N$, let $G/T$ be the quotient of $G \equiv \SpN$
by the right action of $T$. Then the mapping
\begin{equation}\label{eq:polar}
    (G/T) \times \mathfrak{t}_+ \to \mathfrak{E}^0 \;, \quad
    (gT , \Omega) \mapsto g \Omega g^{-1}
\end{equation}
(the reverse of the process of quasi-diagonalization), is a smooth
bijection.

We are stating these facts without proof, as they are standard facts
of symplectic linear algebra.

\section{Probability measure}
\label{sect:III}

By placing a probability distribution on the elliptic domain
$\mathfrak{E}^0$, one gets a random matrix model for disordered
bosonic quasi-particles. We are then interested in the statistics of
the characteristic frequencies or levels $\omega_i\,$.

It is well known that in the Wigner-Dyson situation of random
Hermitian or random real symmetric matrices, where the symmetry group
is compact, the level correlation functions exhibit universal
behavior in a suitable scaling limit. One may therefore ask whether a
similar scenario -- leading to universal laws, possibly of a new kind
-- might be at work in the case being considered.

To answer this question we need to investigate a class of probability
distributions on $\mathfrak{E}^0$ as wide as possible. As a first
step, the present paper deals with a family of well motivated
distributions which are easy to analyze.

\subsection{Choice of measure}

Coming from the standard Wigner-Dyson situation with a compact
symmetry group, one might be inclined to try and consider a Gaussian
distribution
\begin{displaymath}
    P(X)\, dX \stackrel{?}{\propto} \mathrm{e}^{- \Tr\, X^2} dX \;,
\end{displaymath}
where $dX$ is a Lebesgue measure for $\mathfrak{E}^0$:
\begin{equation}
    dX := \prod\nolimits_{i,j} dA_{ij}
    \prod\nolimits_{i \le j} dB_{ij}\, dC_{ij} \;.
\end{equation}
However, such a distribution has infinite mass, since it is invariant
under the action $X \mapsto g X g^{-1}$ by the \textit{non-compact}
group $\SpN$, and it therefore cannot be normalized to be a
probability measure.

Staying within the class of Gaussian distributions, a better choice
of distribution function is
\begin{equation}\label{eq:distr-funct}
    P(X=Jh) \propto \mathrm{e}^{- \tau\, \Tr\, h/2 - \sigma\,
    \Tr\, h^2} = \mathrm{e}^{- \tau\, \Tr\, (J^{-1}X)/2 -
    \sigma\, \Tr\, (J^{-1} X)^2}
\end{equation}
for some positive parameters $\sigma\, , \tau\,$. Because of the
presence of $J^{-1}$ under the trace, this distribution function is
invariant under conjugation $X \mapsto g X g^{-1}$ only if $g \in
\SpN$ satisfies the additional condition $g^{-1} J g = J$. Combining
the two conditions, $g^t J g = J = g^{-1} J g\,$, one sees that the
invariance group of the function $P(X)$ in (\ref{eq:distr-funct}) is
the intersection of the real symplectic and orthogonal groups in $2N$
dimensions:
\begin{equation}\label{eq:def-K}
    K = \SpN \cap \mathrm{SO}_{2N}( \mathbb{R}) \;.
\end{equation}
This group $K$ is isomorphic to $\mathrm{U}_N$, the group of unitary
transformations in $N$ complex dimensions. Indeed, changing from the
symplectic basis $\{ q_1 , \ldots, q_N , p_1 , \ldots, p_N \}$ to the
oscillator basis
\begin{displaymath}
    \{ a_1 , \ldots, a_N , a_1^\dagger , \ldots, a_N^\dagger \} \;,
    \quad a_j = \frac{1}{\sqrt{2}} (q_j + \mathrm{i} p_j) \;, \quad
    a_j^\dagger = \frac{1}{\sqrt{2}} (q_j - \mathrm{i} p_j) \;,
\end{displaymath}
one finds that $K$ is the subgroup of canonical transformations that
do not mix the lowering operators $\{ a_j \}$ with the raising
operators $\{ a_j^\dagger \}$. Moreover, $\mathrm{U}_N \cong K
\subset \SpN$ is known to be a maximal compact subgroup.  It
therefore is the biggest symmetry group possible in our problem.

In the sequel we will consider (\ref{eq:distr-funct}) with $\sigma =
0\;$. Thus we take our probability distribution to be
\begin{equation}\label{eq:meas-0}
    P(X)\,dX := c_N(\tau)\,\mathrm{e}^{-\tau\,\Tr\,(J^{-1}X)/2}dX\,,
\end{equation}
with the normalization constant $c_N(\tau)$ chosen in such a way that
$\int_{\mathfrak{E}^0} P(X)\, dX = 1$. Further motivation for this
choice of distribution was put forth in the introduction (Sect.\
\ref{sect:I}).

\subsection{Polar decomposition and reduction}

Let now $F(X) = F(g X g^{-1})$ be some function on $\mathfrak{E}^0$
which is \textit{radial}, i.e., invariant under conjugation by every
element $g \in \SpN$. Given such a function $F$, which depends only
on the eigenfrequencies $\omega_1, \ldots,\omega_N$ of $X$, we wish
to take the expectation of $F$ w.r.t.\ the probability measure $P(X)
\, dX :$
\begin{equation}\label{eq:averages}
    \left\langle F \right\rangle := \int_{\mathfrak{E}^0} F(X)\, P(X)
    \, dX \;.
\end{equation}
The problem of computing such expectations is best tackled by using
the polar decomposition $\mathfrak{E}^0 \cong \mathfrak{t}_+ \times
(G/T)$ which is given by quasi-diagonalization of $X$; see
(\ref{eq:polar}). Inserting that decomposition into
(\ref{eq:averages}) one has
\begin{equation}\label{eq:reduce}
    \left\langle F \right\rangle = \int_{\mathfrak{t}_+} \left(
    \int_{G/T} P(g \Omega g^{-1})\, dg_T \right) F(\Omega)\, j(\Omega)
    \, d\Omega \;, \qquad d\Omega = d\omega_1 d\omega_2 \cdots
    d\omega_N \;,
\end{equation}
where $g_T$ is a $G$-invariant measure for $G/T$, and $j(\Omega)$ is
the Jacobian of the change of variables $X = g \Omega g^{-1}$.

Let us calculate this Jacobian.  Differentiating the polar coordinate
mapping (\ref{eq:polar}) we get
\begin{displaymath}
    \delta( g \Omega g^{-1} ) = g \left( \delta\Omega +
    [ \, g^{-1}\delta g \, , \Omega ] \right) g^{-1} \;.
\end{displaymath}
The Jacobian we are seeking is the product of all nonzero eigenvalues
of the linear operator $X \mapsto [X,\Omega]$.  These eigenvalues are
called the roots of the pair $(\spN , \mathfrak{t})$.  They are
\begin{displaymath}
    \pm (\omega_i + \omega_j) \quad (i \le j) \;, \quad
    \pm (\omega_i - \omega_j) \quad (i < j) \;,
\end{displaymath}
each with multiplicity one. Thus, by taking the product of all
non-vanishing roots,
\begin{equation}
    j(\Omega)\, d\Omega = \prod_{i < j} (\omega_i^2 - \omega_j^2)^2
    \prod_{k=1}^N (2\omega_k)^2 \,d\omega_k \;.
\end{equation}

To complete the polar integration formula (\ref{eq:reduce}) we need
$\int_{G/T} P(g \Omega g^{-1})\, dg_T$.  In the next subsection we
are going to show that this integral can be calculated in closed form
and depends on $\omega_1 \, \ldots, \omega_N$ as
\begin{equation}\label{eq:HCIZ}
    \int_{G/T} P(g \Omega g^{-1})\, dg_T \propto
    \prod_{i < j} (\omega_i + \omega_j)^{-1}\prod_{k = 1}^N
    \omega_k^{-1} \mathrm{e}^{- \tau \omega_k} \;.
\end{equation}
Thus, in total, the expectation of a radial observable $F(X) =
F(\Omega) \equiv F(\omega_1 , \ldots, \omega_N)$ becomes
\begin{equation}
    \left\langle F \right\rangle = \tilde{c}_N(\tau)
    \int_{\mathbb{R}_+^N} F(\omega_1 , \ldots, \omega_N)
    \prod_{i < j} (\omega_i - \omega_j) (\omega_i^2 - \omega_j^2)
    \prod_{k=1}^N \mathrm{e}^{- \tau \omega_k} \omega_k \, d\omega_k \;,
\end{equation}
where $\tilde{c}_N(\tau)$ is another normalization constant.  This
expectation, for the special choices of $F$ that give the level
correlation functions, will be calculated in Sect.\ \ref{sect:exact}
of the paper.

\subsection{Computation of the integral (\ref{eq:HCIZ})}
\label{Comp of integral}

We now establish (\ref{eq:HCIZ}). Omitting a normalization constant,
we denote the integral on the left-hand side of (\ref{eq:HCIZ}) by
\begin{equation}
    I(\Omega) := \int_{G/T} \mathrm{e}^{-\tau\,
    \Tr \, (J^{-1} g \Omega g^{-1})/2} dg_T \;.
\end{equation}
What makes this integral computable in closed form is that $J$ lies
in $\spN$ and $\Omega \mapsto g \Omega g^{-1}$ is the adjoint action
of $G = \SpN$ on its Lie algebra. These circumstances place the
integral in the class of integrals of Harish-Chandra-Itzykson-Zuber
type, which are covered by the Duistermaat-Heckman theorem and its
generalizations. In the present case, the integral can be computed in
a particularly simple manner, as follows.

Let $dg$ and $dt$ be Haar measures for $G$ and $T$, respectively,
with $dg = dg_T \, dt$ and $\int_T dt = \mathrm{vol}(T)$. Our first
step is to switch from $G/T$ to integrating over the full symplectic
group $G:$
\begin{displaymath}
    I(\Omega) = \frac{1}{\mathrm{vol}(T)} \int_G \mathrm{e}^{-
    \tau\, \Tr \, (J^{-1} g \Omega g^{-1})/2}\, dg \;.
\end{displaymath}

Next we use that $dg$ is invariant under inversion, $g \mapsto
g^{-1}$. After this transformation the integrand is expressed in
terms of the combination $g J^{-1} g^{-1} = - g J g^{-1}$. Since $k J
k^{-1} = J$ for $k \in K \cong \mathrm{U}_N\,$, we can push down the
resulting integral over $G$ to an integral over the quotient space
$G/K$. Let $dg_K$ and $dk$ be invariant resp.\ Haar measures for
$G/K$ and $K$ so that $dg = dg_K \, dk\,$. Then
\begin{equation}\label{eq:I(Omega)}
    I(\Omega) = \frac{\mathrm{vol}(K)}{\mathrm{vol}(T)} \int_{G/K}
    \mathrm{e}^{\tau\, \Tr \, (\Omega g J g^{-1})/2}\, dg_K \;,
    \quad \mathrm{vol}(K) = \int_K dk \;.
\end{equation}

The homogeneous space $G/K \cong \SpN / \mathrm{U}_N$ has the salient
feature of being a non-compact symmetric space of Hermitian type.
Such spaces carry the structure of a K\"ahler manifold, which means
that $G/K$ comes with a non-degenerate, closed, and $G$-invariant
2-form (the K\"ahler form of $G/K$). Writing $g J g^{-1} =: Q$ this
is the form
\begin{equation}
    \beta = \Tr\, (Q \, \mathrm{d}Q \wedge \mathrm{d}Q) \;.
\end{equation}
Notice that $\mathrm{dim}_\mathbb{R}\, G/K = N(2N+1) - N^2 = N(N+1)$.
Raising $\beta$ to its $\frac{1}{2}N(N+1)^\mathrm{th}$ exterior power
one obtains a top-dimensional form, $\beta^{\frac{1}{2}N(N+1)}$,
which is still $G$-invariant and nonzero. Since $G/K$ is homogeneous,
there can be at most one such form up to multiplication by scalars.
Therefore, there exists some (nonzero) constant such that
\begin{equation}\label{eq:inv-meas}
    dg_K = \mathrm{const} \times \beta^{\frac{1}{2}N(N+1)} \;.
\end{equation}

By Darboux's theorem one can find local symplectic coordinates for
$G/K$ that bring $\beta$ into canonical form. While this fact by
itself would not be of much practical help, in the present case such
coordinates exist \textit{globally} and, moreover, they can be chosen
in such a way that $\Tr (\Omega\, g J g^{-1})$ depends on them
\textit{quadratically}.

To describe these perfect coordinates, consider the space of complex
symmetric $N \times N$ matrices, $\mathrm{Sym} (\mathbb{C}^N)$, which
has dimension $\frac{1}{2} N (N+1)$ over $\mathbb{C}$ and thus shares
with $G/K$ the dimension $N(N+1)$ over $\mathbb{R}\,$. With every $Z
\in \mathrm{Sym}( \mathbb{C}^N )$ associate a positive Hermitian $2N
\times 2N$ matrix $\widetilde{g}$ by
\begin{equation}\label{eq:g-tilde}
    \widetilde{g} \equiv \widetilde{g}(Z,Z^\dagger) =
    \begin{pmatrix} (1 + Z Z^\dagger)^{1/2} &Z\\
    Z^\dagger &(1 + Z^\dagger Z)^{1/2} \end{pmatrix} \;.
\end{equation}

Now if $S$ is the matrix of the unitary transformation from the real
symplectic basis $\{ p_j \, , q_j \}$ of $\mathbb{R}^{2N}$ to the
oscillator basis $\{ a_j \, , a_j^\dagger \}$:
\begin{displaymath}
    S := \frac{1}{\sqrt{2}} \begin{pmatrix} 1_N &
    \mathrm{i}\, 1_N\\ - 1_N &\mathrm{i} \, 1_N \end{pmatrix} \;,
\end{displaymath}
then $g := S^{-1} \widetilde{g} S$ is immediately seen to be a real
matrix and, using the relation
\begin{displaymath}
    S J S^{-1} = \mathrm{i} \Sigma_3 \;, \quad \Sigma_3 =
    \begin{pmatrix} 1_N &0\\ 0 &-1_N \end{pmatrix} \;,
\end{displaymath}
one finds that $g = S^{-1} \widetilde{g} S$ satisfies $g^\dagger J g
= g^t J g = J$ and hence lies in $\SpN$. Moreover, the reverse
correspondence $k \mapsto \widetilde{k} = S k S^{-1}$ is the
isomorphism between $K$ and $\mathrm{U}_N$ discussed in the paragraph
after (\ref{eq:def-K}); it takes $k \in K$ to the block-diagonal form
\begin{displaymath}
    \widetilde{k} = \begin{pmatrix} U &0\\ 0 &\bar{U}
    \end{pmatrix} \;, \quad U \in \mathrm{U}_N \;.
\end{displaymath}
It is now clear that the mapping $\mathrm{Sym}( \mathbb{C}^N ) \to
G/K$ by $Z \mapsto S^{-1} \widetilde{g}(Z,Z^\dagger) S \cdot K \equiv
gK$ is a bijection. Using it to express the K\"ahler form $\beta$ in
terms of the complex symmetric matrix $Z$, one obtains
\begin{equation}
    \beta = \Tr\, (Q\, \mathrm{d}Q \wedge \mathrm{d}Q) = - 4
    \mathrm{i}\, \Tr\, (\Sigma_3 \, \mathrm{d} \widetilde{g}^{-1}
    \wedge \mathrm{d}\widetilde{g}) = 8\mathrm{i}\, \Tr \,
    (\mathrm{d}Z \wedge \mathrm{d}Z^\dagger) \;.
\end{equation}
Thus, the top-dimensional form $\beta^{\frac{1}{2}N(N+1)}$ is
constant in $Z:$
\begin{equation}
    \frac{\beta^{\frac{1}{2}N(N+1)}}{(\frac{1}{2}N(N+1))!} =
    (8\mathrm{i})^{\frac{1}{2} N(N+1)} 2^{\frac{1}{2} N(N-1)}
    \prod\nolimits_{i \le j} dZ_{ij} \wedge d\bar{Z}_{ij} \;,
\end{equation}
and from (\ref{eq:inv-meas}) the invariant measure $dg_K$ is a
constant multiple of the Lebesgue measure for $\mathrm{Sym}
(\mathbb{C}^N)$.

Finally, from $g = S^{-1} \widetilde{g} S\,$, $S J S^{-1} =
\mathrm{i} \Sigma_3$ and (\ref{eq:g-tilde}) one has
\begin{displaymath}
    - \Tr \,(\Omega\, g J g^{-1}) = \Tr\, \omega (1 + 2 Z Z^\dagger)
    + \Tr \, \omega (1 + 2 Z^\dagger Z) \;, \quad \omega =
    \mathrm{diag}(\omega_1 , \ldots, \omega_N) \;.
\end{displaymath}
Our integral (\ref{eq:I(Omega)}) now becomes a Gaussian integral:
\begin{displaymath}
    I(\Omega) = \mathrm{const} \times \mathrm{e}^{-\tau \sum_k \omega_k}
    \int \mathrm{e}^{-2 \tau \Tr \, Z^\dagger (Z \omega + \omega Z)}
    \prod\nolimits_{i \le j} dZ_{ij} \, d\bar{Z}_{ij} \;.
\end{displaymath}
Doing this integral one immediately obtains the result for
$I(\Omega)$ stated in (\ref{eq:HCIZ}).

\subsection{Generalization}

A slight generalization of (\ref{eq:meas-0}) is afforded by the
observation that the determinant of $X$ in (\ref{eq:sympl-Lie}) is
always positive:
\begin{displaymath}
    \mathrm{Det}(X)=\mathrm{Det}(\Omega)=\prod_{k=1}^N \omega_k^2\;.
\end{displaymath}
Thus, by multiplying the probability measure $P(X)\, dX$ by some
power $l - 1 > - 1$ of the positive square root $\mathrm{Det}
(X)^{1/2}$ and adjusting the normalization constant, we get another
probability measure:
\begin{equation}
    P_l(X)\, dX = \mathrm{const} \times \mathrm{Det}(X)^{\frac{1}{2}
    (l-1)}\, \mathrm{e}^{-\frac{1}{2} \tau\, \mathrm{Tr}\,(J^{-1} X)}dX\;.
\end{equation}
This measure is still $\mathrm{U}_N$-invariant. By the process of
quasi-diagonalization and drawing on our results above, we push it
forward to a measure for the eigenfrequencies.  The result is
\begin{equation}\label{eq:ell-meas}
    d\mu_{N,l}(\omega_1, \ldots, \omega_N) = c_{N,l}(\tau)
    \prod\limits_{i < j} (\omega_i - \omega_j)^2 (\omega_i + \omega_j)
    \prod\limits_{k =1}^N \omega_k^l \, \mathrm{e}^{-\tau \omega_k}\,
    d\omega_k \;.
\end{equation}
This, for any non-negative power $l \in \mathbb{Z}\,$, is the family
of probability distributions to be studied in the present paper.

\section{Large-$N$ limit of the 1-point density in the bulk}
\label{sect:cubic}

The 1-point density $\rho(\omega)\, d\omega$ is defined as the
probability density for any one of the eigenfrequencies $\omega_i$ to
have the value of $\omega\,$, irrespective of what the values of the
other eigenfrequencies are; thus $\rho(\omega)$ is the function
\begin{equation}
    \rho(\omega) := \int \sum_{i = 1}^N \delta(\omega - \omega_i) \,
    d\mu_{N,l}(\omega_1 , \ldots, \omega_N) \;,
\end{equation}
which has the properties $\rho(\omega) \ge 0$ and
\begin{equation}\label{eq:norm}
   \int_0^\infty \rho(\omega) \, d\omega = N \;.
\end{equation}

We are now interested in the behavior of the density function $\rho
(\omega)$ in the limit of $N \to \infty$.  From the expression
(\ref{eq:ell-meas}) and experience with similar problems (see, e.g.,
\cite{mehta}), we expect that this limit can be obtained by maximizing
the functional
\begin{equation}\label{eq:functional}
    F = \frac{1}{2} \int_0^\infty \int_0^\infty \ln \big( (\omega
    - \omega^\prime )^2 (\omega + \omega^\prime) \big) \,
    \rho(\omega) \rho(\omega^\prime) \, d\omega^\prime d\omega +
    \int_0^\infty \ln ( \omega^l\, \mathrm{e}^{- \omega \tau})\,
    \rho(\omega) \, d\omega
\end{equation}
subject to the constraint (\ref{eq:norm}) and the condition
$\rho(\omega) \ge 0$. More precisely, the limit is expected to exist
in the scaled variable $x := \omega \tau /N$; i.e., there should
exist a certain non-negative function $\rho_\infty(x)$ with $\int
\rho_\infty(x) \, dx = 1$ such that $\rho(\omega)$ is asymptotic to
$\tau \rho_\infty(\omega \tau / N)$.

Varying $F$ with respect to $\rho(\omega)$ we get
\begin{displaymath}
    \frac{\delta F}{\delta\rho(\omega)} = \int_0^\infty \big(2\ln|\omega
    -\omega^\prime| +\ln(\omega+\omega^\prime)\big)\,\rho(\omega^\prime)
    \, d\omega^\prime + l\, \ln\omega - \omega \tau \;.
\end{displaymath}
We now insert the asymptotic equality $\rho(Nx/\tau) \approx \tau
\rho_\infty (x)$ and pass to the limit $N \to \infty$ in the scaling
variable $x\,$. Let $\mathrm{supp}(\rho_\infty) = [0,b]$ be the
region of support of $\rho_\infty\,$. Then the condition $\delta F /
\delta \rho(\omega) = N \lambda\,$, where $\lambda$ is a Lagrange
multiplier for the constraint (\ref{eq:norm}), yields the equation
\begin{equation}\label{eq:before-diff}
    \int_0^b \big( 2\ln|x - x^\prime| + \ln(x + x^\prime)\big)\,
    \rho_\infty(x^\prime)\, dx^\prime - x = \lambda \qquad (0<x<b)\;,
\end{equation}
which no longer depends on the parameter $l\,$. It can be shown that
our functional $F$ is convex; as a result, the solution $\rho_\infty$
of Eq.\ (\ref{eq:before-diff}) exists and is unique when supplemented
by the normalization condition
\begin{equation}\label{eq:4.5}
    \int_0^b \rho_\infty(x) \, dx = 1 \;.
\end{equation}

In the following subsections, we are going to construct the solution
to the mathematical problem posed by (\ref{eq:before-diff}) and
(\ref{eq:4.5}). It will turn out to be
\begin{equation}\label{eq:soln-rho}
    \rho_\infty(x) = \frac{1}{2\pi} (x/b)^{-1/3}\big( (1 +
    \sqrt{1-x^2/ b^2})^{1/3} - (1 - \sqrt{1-x^2/b^2})^{1/3}\big)
    \qquad (0 < x \le b = 3 \sqrt{3})\;.
\end{equation}
\begin{figure}
        \parbox[l]{8.5cm}{\input{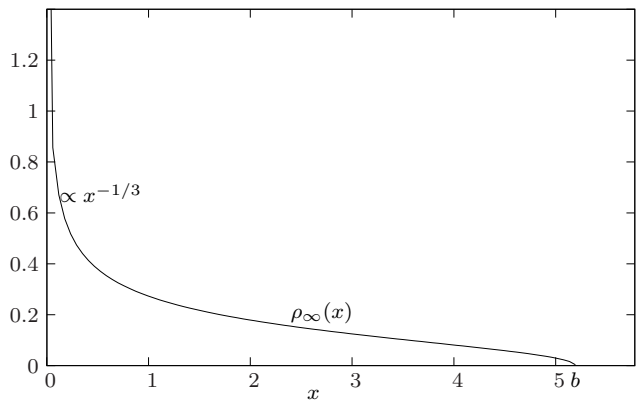}}
        \caption{The graph of the function $\rho_\infty\,$.}
        \label{fig:1}
\end{figure}
The graph of this function is plotted in Fig.\ \ref{fig:1}. From the
expression (\ref{eq:soln-rho}) the behavior near the lower edge $x =
0$ is
\begin{displaymath}
   \rho_\infty(x) \simeq \frac{1}{2\pi} (2b/x)^{1/3} \qquad
   (0 < x \ll b) \;,
\end{displaymath}
while close to the upper edge $x = b$ one gets
\begin{displaymath}
   \rho_\infty(x) \simeq \frac{1}{3\pi} (1 - x^2 /b^2)^{1/2}
   \qquad (x < b\, , \,\, x \to b)\;.
\end{displaymath}

In the vicinity of the upper and lower edges there exists crossover
to a fine-scale behavior that cannot be found by the present method
of maximization of the functional $F$. The crossover at the upper
edge involves Airy functions on a scale $N^{1/3}$, which is small
compared to the bulk scale $N$. At the lower edge, the crossover
occurs on a very fine scale, $N^{-1/2}$, which is small even in
comparison with the bulk mean level spacing (which is of order
$N^0$).

\subsection{Method of solution (idea)}

We do not know how to solve Eq.\ (\ref{eq:before-diff}) for the
unknown function $\rho_\infty(x)$ directly. Therefore, to simplify
the problem we differentiate once with respect to $x$ to obtain the
equation
\begin{equation}\label{eq:Hilbert}
    2{\cal P} \int_0^b \frac{\rho_\infty(x')\, dx^\prime}{x-x'} +
    \int_0^b \frac{\rho_\infty(x')\, dx^\prime}{x + x'} = 1 \;,
\end{equation}
where ${\cal P}$ means the principal value of the integral. At this
stage, the value of $b$ is unknown but assumed to be finite.

Introducing the Green's function (or Stieltjes transform)
\begin{equation}\label{eq:Green}
    g(z) := \int_0^b \frac{\rho(x) \, dx}{x - z} \;, \quad
    z \in \mathbb{C}\setminus [0,b] \;,
\end{equation}
and the related functions
\begin{equation}
    g_\pm(x) := \lim_{\varepsilon \to 0+} g(x \pm
    \mathrm{i}\varepsilon) \;, \qquad g_0(x) := - g(-x) \;,
\end{equation}
we bring Eq.\ (\ref{eq:Hilbert}) into the form
\begin{equation}\label{eq:ReG}
   g_+(x) + g_-(x) + g_0(x) = - 1 \qquad (0 < x < b) \;.
\end{equation}
To solve this equation, we are led to do an exercise in complex
analysis which is motivated as follows.

Let $w \mapsto f(w)$ be some meromorphic function of a complex
variable $w$, and let the equation $z = f(w)$ have $r$ simple roots
$w_1(z)\, , w_2(z)\, ,\ldots, w_r(z)$, i.e., $z = f(w_1(z)) = \ldots
= f(w_r(z))$. If the function $f$ is analytic in $1/w$ at $w =
\infty$, these roots add up to a constant:
\begin{equation}\label{eq:root-sum}
    \sum_{i=1}^r w_i(z) = \mathrm{const} =: c \quad \text{(independent of
    $z$)}\;.
\end{equation}
Indeed, if $\gamma$ is a closed contour encircling all of the roots
in the counterclockwise sense, then
\begin{displaymath}
    \sum w_i^\prime(z) = \sum \frac{1} {f^\prime(w_i(z))} =
    \frac{1}{2\pi\mathrm{i}} \oint_\gamma \frac{dw}{f(w) - z} = 0 \;,
\end{displaymath}
where the second equality is by the residue theorem, and the last
equality follows by contracting $\gamma$ to the point at infinity.
Thus $\sum w_i^\prime (z) = 0$ and hence $\sum w_i(z) =
\mathrm{const}$.

Eq.\ (\ref{eq:root-sum}) for $r = 3$ looks similar to (\ref{eq:ReG})
and can, in fact, be made to look identical to it by the following
observation. Notice that the function $z \mapsto g(z)$ defined by
(\ref{eq:Green}) is holomorphic in the interior of the left half of
the complex plane. Suppose, therefore, that we have found a root
$g(z)$ of $z = f(g(z))$ which is holomorphic in the left half plane,
and that $g_\pm (x) = \lim_{\varepsilon \to 0+} g(x \pm \mathrm{i}
\varepsilon)$ are its two analytic continuations to positive real $x
\in (0,b)$. Moreover, suppose that the function $f$ has a reflection
symmetry
\begin{equation}\label{eq:refl-sym}
    f(w) = - f(2a-w) \qquad (a \in \mathbb{C}) \;.
\end{equation}
Then $z \mapsto 2a - g(-z)$ is a root of $z = f(w)$ holomorphic in
the right half plane, and from (\ref{eq:root-sum}) we infer that
\begin{displaymath}
    g_+(x) + g_-(x) + \big( 2a - g(-x) \big) = c \;.
\end{displaymath}
Setting $g_0(x) \equiv - g(-x)$ this becomes the same as
(\ref{eq:ReG}) if
\begin{equation}\label{eq:const}
    c - 2a = -1 \;.
\end{equation}

Thus we are inspired to interpret $g_+\,$, $g_-\,$, and $2a + g_0$ as
the three roots of an equation $z = f(w)$. Given this interpretation,
solving (\ref{eq:ReG}) amounts to finding the function $f$.

\subsection{The good function $f$ to consider}

We are looking for a certain meromorphic function $f$ on $\mathbb{C}
\,$. By adding a point at infinity we can view such a function $f$ as
a mapping of the Riemann sphere $\mathrm{S}^2 = \mathbb{C} \cup \{
\infty \}$ to itself. We want this mapping to have degree $r = 3\,$;
i.e., every regular point $z$ of $f$ is to have three distinct
pre-images:
\begin{displaymath}
    f^{-1}(z) = \{ w_1(z) , w_2(z) , w_3(z) \} \;.
\end{displaymath}
Such a mapping can be presented in the general form
\begin{equation}\label{eq:f(w)}
    f(w) = f_\infty + \sum_{i=1}^3 \frac{b_i}{w - a_i}
\end{equation}
with some complex numbers $a_i\,$, $b_i\,$, and $f_\infty\,$.

Let us narrow down the choice of parameters. From the normalization
condition (\ref{eq:4.5}) and the definition of $g(z)$ in
(\ref{eq:Green}), we have the limit $z\, g(z) \to -1$ for $z \to
\infty\,$. Therefore, since $f(g(z)) = z$ by construction, we need
$f(w)$ to have a pole at $w = g(\infty) = 0$ with residue $-1$. So we
choose $a_1 = 0$ and $b_1 = -1$. The reflection symmetry
(\ref{eq:refl-sym}) is then implemented by setting $f_\infty = 0\,$,
$b_3 = b_1\,$, and $a_i = (i-1)\,a$ for $i = 1, 2, 3$ and some $a \in
\mathbb{C}\,$. Thus,
\begin{displaymath}
    f(w) = - \frac{1}{w} + \frac{b_2}{w-a} - \frac{1}{w-2a} \;,
\end{displaymath}
where the parameters $a$ and $b_2$ are still unknown.

Next observe that for a degree-$r$ holomorphic mapping $f : \,
\mathrm{S}^2 \to \mathrm{S}^2$, the number of singular points, where
$f^\prime(w) = 0\,$, is $2r-2\,$. Indeed, writing $f$ as $f(w) =
p(w)/q(w)$ where $p$ and $q$ are polynomials of degree $r\,$, one has
\begin{displaymath}
    f^\prime(w) = \frac{p^\prime(w)\, q(w) - p(w) \, q^\prime(w)}
    {q(w)^2} \;,
\end{displaymath}
the numerator of which is a polynomial of degree $2r-2$ and so has
$2r - 2$ zeroes.

Thus we should expect our function (\ref{eq:f(w)}) to have $2 \times
3 - 2 = 4$ singular points. The reflection symmetry
(\ref{eq:refl-sym}) makes for their images $\{ f(w) \in \mathbb{C} \,
|\, f^\prime(w) = 0 \}$ to be arranged symmetrically around $z = 0$.
Now notice that our Green's function $g(z)$, being the Stieltjes
transform of $\rho_\infty(x)$ with support $[0,b]\,$, must have
singularities at $z = 0$ and $z = b\,$. The image of the singular set
had better contain these values, and thus is determined to be $\{-b\,
, 0\, , +b\}$ by reflection symmetry. Actually, since our situation
calls for $f$ to have four singular points, the singularity at $z =
0$ (corresponding to $w = \infty$) must have multiplicity two. This
is achieved by choosing $b_2 = - b_1 - b_3 = +2\,$, so that
\begin{displaymath}
    f(w) = - \frac{1}{w} + \frac{2}{w-a} - \frac{1}{w-2a} =
    \frac{-2a^2}{w(w-a)(w-2a)} \;,
\end{displaymath}
resulting in the behavior $f(w) \sim w^{-3}$ for $w \to \infty$. The
singular points of $f$ now are $w = a \pm a / \sqrt{3}\,$, and
$\infty$. These correspond to $z = f(w) = \pm 3 \sqrt{3} / a\,$, and
$0\,$, respectively, so we infer
\begin{equation}\label{eq:b-from-a}
    b = 3 \sqrt{3} / |a| \;.
\end{equation}

It remains to pin down the last unknown parameter $a\,$. For that
purpose, recall that the sum of the roots $f^{-1}(z) = \{ w_1(z) ,
w_2(z) , w_3(z) \}$ is a constant, $c\,$, independent of $z\,$. To
determine this constant, look at $\sum w_i(\infty)$ and use that the
poles of $f$ are at $w = 0\,$, $a\,$, $2a$ to obtain
\begin{equation}\label{eq:sum-const}
    c = \sum w_i(z) = \sum w_i(\infty) = 3a \;.
\end{equation}
We then conclude $a = -1$ from (\ref{eq:const}), and $b = 3\sqrt{3}$
from (\ref{eq:b-from-a}). In summary, the good meromorphic function
$f$ for us to consider is
\begin{equation}\label{eq:cubic}
    w \mapsto f(w) = \frac{-2}{w(w+1)(w+2)} \;.
\end{equation}
Let us mention in passing that the idea to consider the equation $z =
f(w)$ or, equivalently,
\begin{displaymath}
    w(w+1)(w+2) + 2/z = 0 \;,
\end{displaymath}
first came to one of us (HJS) from previous work \cite{sz} on the
Green's function of the Bures measure, whose large-$N$ limit leads to
a similar equation.

\subsection{Solution of the problem}

The situation can now be succinctly described like this: thinking of
\begin{displaymath}
    W:= \mathbb{C} \setminus \{ -1 + 1/\sqrt{3}\, , -1 -1 / \sqrt{3} \}
    \;, \qquad Z := \big( \mathbb{C} \setminus \{ b\, , 0\,, -b \} \big)
    \cup \{ \infty \} \quad (b = 3\sqrt{3}) \;,
\end{displaymath}
as two Riemann surfaces $W$ and $Z$, the function $f$ of
(\ref{eq:cubic}) gives us a holomorphic cover
\begin{displaymath}
    f: \, W \to Z \;, \qquad f^{-1}(z) = \{w_1(z),w_2(z),w_3(z)\}\;.
\end{displaymath}

What's the monodromy of this cover, i.e., what happens when the
locally defined functions $z \mapsto w_i(z)$ are analytically
continued around one of the singular points $z = b\, , 0\, , -b\,$?
At the point $z = 0$ (or $w = \infty$) we have a cubic singularity $z
\sim w^{-3}$. Consequently, the monodromy at $z = 0$ cyclically
permutes the roots $w_i(z)$.  Turning to $z = \pm b\, $, we see that
linearization $z = \pm b + \delta z$ and $w = f^{-1}(\pm b) + \delta
w$ gives
\begin{displaymath}
    \delta z \sim (\delta w)^2 \;.
\end{displaymath}
In the latter two cases the monodromy must exchange two of the
$w_i(z)$ while leaving the third one invariant.

Now focus on the situation near the singular point $z = - b$ and
denote by $w(z) \equiv g(z)$ the root which, there, is trivial under
monodromy and hence exists as a holomorphic function in some
neighborhood of $z = - b\,$. With the remaining two singularities
being at $z = 0$ and $z = b\,$, the function $g(z)$ actually extends
to a holomorphic function on the Riemann sphere $\mathbb{C} \cup \{
\infty \}$ cut along, say, $[0,b] \subset \mathbb{R}\,$. Let us
verify that this holomorphic function $g : \, (\mathbb {C} \setminus
[0,b]) \cup \{ \infty \} \to W$ coincides with the Green's function
(\ref{eq:Green}) solving our problem (\ref{eq:ReG}).

By the holomorphic nature of $g$ and Cauchy's theorem, we have that
\begin{displaymath}
    g(z) = \frac{1}{2\pi\mathrm{i}} \oint_\gamma
    \frac{g(z^\prime)\, dz^\prime}{z^\prime - z} \;,
\end{displaymath}
where $\gamma$ is a small loop running around $z$ in the
counterclockwise sense. Since $g$ is holomorphic at infinity, the
loop $\gamma$ can be deformed (through infinity) to a loop encircling
the cut $[0,b]\,$, but now with the orientation reversed. Collapse
the deformed loop to the two line segments connecting $0$ with $b\,$.
Then, setting $g_\pm(x) = \lim_{\varepsilon\to 0+} g(x \pm \mathrm{i}
\varepsilon)$ and
\begin{equation}\label{eq:jump}
    \rho_\infty(x) := \frac{g_+(x) - g_-(x)}{2\pi\mathrm{i}}
    \qquad (0 < x < b) \;,
\end{equation}
$g(z)$ is obviously given by the integral in (\ref{eq:Green}).

Because $g_+(x)$ and $g_-(x)$ arise by analytic continuation from
$g(z) \in f^{-1}(z)$, these are two of the three elements in the set
$f^{-1}(x)$. How is the third element of $f^{-1}(x)$ related to
$g(z)$? To see that, recall $a = -1$ and from (\ref{eq:refl-sym}) the
invariance of the equation $z = f(w)$ under $(z,w) \mapsto
(-z,2a-w)$. Thus, if $g(-z)$ is a root over $-z\,$, then $-2-g(-z)$
is a root over $z\,$, and it follows that $g_0(x)-2$ with $g_0(x) :=
- g(-x)$ (for $0 < x < b$) is a root over $x\,$. The roots $g_+(x)$,
$g_-(x)$, and $g_0(x)-2$ all are different as functions. In fact,
$\mathfrak{Im}\, g_+(x) > 0 = \mathfrak{Im}\, g_0(x) >
\mathfrak{Im}\, g_-(x)$ for $0 < x < b\,$.  So,
\begin{displaymath}
    f^{-1}(x) = \{ g_+(x) \, , \, g_-(x)\, ,\, g_0(x)-2 \} \;,
\end{displaymath}
and from (\ref{eq:sum-const}) we deduce that
\begin{displaymath}
    g_+(x) + g_-(x) + g_0(x) - 2 = \sum_{w \in f^{-1}(x)} w = c = 3a = -3
    \qquad (0 < x < b)\;,
\end{displaymath}
which agrees with (\ref{eq:ReG}). Recall that in order for our
analysis to work out we must choose
\begin{equation}\label{eq:b-value}
    b = 3 \sqrt{3} \;.
\end{equation}

With a full understanding of the situation in hand, it is now an easy
exercise to obtain $\rho_\infty(x)$ in explicit form. Solving the
equation $z = f(w)$ one finds the holomorphic function $g(z)$ in the
interval $-b < x < 0$ to be
\begin{displaymath}
    g(x) = (-x)^{-1/3} (1 + \sqrt{1 - x^2/b^2})^{1/3} + (-x)^{-1/3}
    (1 - \sqrt{1 - x^2/b^2})^{1/3} - 1 \;,
\end{displaymath}
where all square roots and cubic roots are understood to be positive.
This function indeed extends holomorphically to a neighborhood of $x
= -b\,$, as the Taylor expansion at $x = -b$ contains only even
powers of $\sqrt{1 - x^2/b^2}$. Analytic continuation around the
singularity at $z = 0$ gives
\begin{displaymath}
    g_\pm(x) = x^{-1/3} \mathrm{e}^{\pm \pi\mathrm{i}/3} (1 +
    \sqrt{1 - x^2/b^2})^{1/3} + x^{-1/3} \mathrm{e}^{\mp \pi
    \mathrm{i}/3} (1 - \sqrt{1 - x^2/b^2})^{1/3} - 1 \qquad
    (0 < x \le b) \;.
\end{displaymath}
Computing the difference (\ref{eq:jump}) we then get the result for
$\rho_\infty(x)$ claimed in (\ref{eq:soln-rho}), with the value for
$b$ given by (\ref{eq:b-value}).

As a final remark, let us note that the good form of $g(z)$ to use
near infinity is
\begin{equation}
    g(z) = -1 +
    \mathrm{e}^{\mathrm{i} \pi / 6} \left(\sqrt{\frac{1}{b^2} -
    \frac{1}{z^2}} + \frac{\mathrm{i}}{z} \right)^{1/3} +
    \mathrm{e}^{-\mathrm{i} \pi / 6} \left(\sqrt{\frac{1}{b^2} -
    \frac{1}{z^2}} - \frac{\mathrm{i}}{z} \right)^{1/3} \;.
\end{equation}
From this, all moments of $\rho_\infty(x)\, dx$ can be found by
expanding $g(z)$ in powers of $1/z$.

\section{Exact solution using bi-orthogonal polynomials}
\label{sect:exact}

We now express the probability measure (\ref{eq:ell-meas}) as
\begin{equation}\label{eq:new-meas}
    d\mu_{N,l}(\omega_1, \ldots, \omega_N) = c_{N,l}(\tau)
    \prod\limits_{i < j} (\omega_i - \omega_j) (\omega_i^2 -
    \omega_j^2) \prod\limits_{k=1}^N \mathrm{e}^{-\tau\omega_k}
    \,\omega_k^l \,d\omega_k \;,
\end{equation}
and embark on another approach to handling it.

To get started, recall the formula for the Vandermonde determinant:
\begin{displaymath}
    \prod_{i > j} (\omega_i - \omega_j) = \mathrm{Det}\,
    (\omega_j^{i-1})_{i,j=1,\ldots,N} = \left\vert \begin{matrix}
    1 &1 &\ldots &1 \\ \omega_1 &\omega_2 &\ldots &\omega_N \\
    \vdots &\vdots &\ddots &\vdots\\ \omega_1^{N-1} &\omega_2^{N-1}
    &\ldots &\omega_N^{N-1} \end{matrix} \right\vert \;.
\end{displaymath}
Using it we reorganize the probability measure (\ref{eq:new-meas}) as
\begin{equation}\label{eq:new1-meas}
    d\mu_{N,l}(\omega_1, \ldots, \omega_N) = c_{N,l}(\tau)\,\mathrm{Det}\,
    (\omega_j^{i-1})\,\mathrm{Det}\,(\omega_j^{2i-2})\prod\limits_{k=1}^N
    \mathrm{e}^{-\tau\omega_k}\,\omega_k^l \,d\omega_k \;.
\end{equation}
We also simplify our notation by setting $\tau = 1$.

By standard properties of the determinant, $\mathrm{Det}\, (\omega_j^
{i - 1})$ changes only by a multiplicative constant when the
monomials $\omega_j^{i-1}$ are replaced by any polynomials in
$\omega_j$ of degree $i-1\,$. We have two Vandermonde determinants,
$\prod_{i<j} (\omega_i - \omega_j)$ and $\prod_{i<j} (\omega_i^2 -
\omega_j^2)$, so we introduce two sets of polynomials, denoting those
of the first set by $P_{i-1}(\omega_j)$ and those of the second one
by $Q_{i-1} (\omega_j^2)$. Our measure then becomes
\begin{equation}\label{eq:new2-meas}
    d\mu_{N,l}(\omega_1,\ldots,\omega_N) = \tilde{c}_{N,l}\, \mathrm{Det}\,
    \big(P_{i-1}(\omega_j)\big)\, \mathrm{Det}\,\big(Q_{i-1}(\omega_j^2)\big)
    \prod\limits_{k=1}^N \mathrm{e}^{-\omega_k}\,\omega_k^l\,d\omega_k\;.
\end{equation}

In order for the introduction of the polynomials $P_n(\omega)$ and
$Q_n (\omega^2)$ to be useful we require them to be orthogonal with
respect to the integration measure $\mathrm{e}^{-\omega} \omega^l \,
d\omega :$
\begin{equation}\label{eq:ortho}
    I_{m,n} \equiv \int_0^\infty P_m(\omega)\, Q_n(\omega^2)\,
    \mathrm{e}^{-\omega}\omega^l\, d\omega = h_n\, \delta_{m,n}\;,
\end{equation}
where the numbers $h_n = I_{n,n}$ depend on the choice of
normalization for $P_n(\omega)$ and $Q_n(\omega^2)$. Such polynomials
are constructed by a variant of the Gram-Schmidt algorithm, as
follows.

\subsection{Bi-orthogonal polynomials}\label{sect:V.A}

We review the construction in the general setting of two real vector
spaces $V, W$ with a pairing (or non-degenerate bilinear form)
\begin{displaymath}
    \gamma : \, V \times W \to \mathbb{R} \;.
\end{displaymath}
Given some basis $v_0, v_1, v_2, \ldots$ of $V$, and a basis $w_0,
w_1, w_2, \ldots$ of $W$, let the entries of the corresponding
pairing matrix be denoted by
\begin{displaymath}
    \gamma_{m,n} := \gamma(v_m\,, w_n) \qquad (m,n = 0,1,2,\ldots)\;.
\end{displaymath}
The goal now is to construct a new basis $e_0, e_1, e_2, \ldots$ of
$V$, and a new basis $f_0 , f_1 , f_2, \ldots$ of $W$ such that
\begin{displaymath}
    e_n = v_n + \sum_{n^\prime = 0}^{n-1} A_{n n^\prime}
    \, v_{n^\prime} \;, \qquad f_n = w_n + \sum_{n^\prime =
    0}^{n-1} B_{n n^\prime} \, w_{n^\prime} \;,
\end{displaymath}
(with real coefficients $A_{n n^\prime}$ and $B_{n n^\prime}$), and
the transformed basis vectors form a bi-orthogonal system:
\begin{displaymath}
    \gamma(e_m \, , f_n) = 0 \qquad m \not= n \;.
\end{displaymath}

This problem has a unique solution by the process of Gram-Schmidt
orthogonalization. A nice way of presenting the solution is by means
of the following determinants (where, by a slight abuse of notation,
the matrix entries in the last column resp.\ last row are vectors,
whereas all of the other matrix entries are numbers):
\begin{equation}\label{eq:Gram-Schmidt}
    e_n = D_{n-1}^{-1}\, \left\vert \begin{matrix} \gamma_{0,0}
    &\ldots &\gamma_{0,n-1} &v_0 \\ \gamma_{1,0} &\ldots &\gamma_{1,n-1}
    &v_1 \\ \vdots &\ddots &\vdots &\vdots \\ \gamma_{n,0} &\ldots
    &\gamma_{n,n-1} &v_n \end{matrix} \right\vert \;, \qquad
    f_n = D_{n-1}^{-1}\, \left\vert \begin{matrix} \gamma_{0,0}
    &\gamma_{0,1} &\ldots &\gamma_{0,n}\\ \vdots &\vdots &\ddots
    &\vdots \\ \gamma_{n-1,0} &\gamma_{n-1,1} &\ldots &\gamma_{n-1,n}\\
    w_0 &w_1 &\ldots &w_n \end{matrix} \right\vert \qquad (n = 1, 2,
    \ldots),
\end{equation}
with normalization factor
\begin{displaymath}
    D_n = \left\vert \begin{matrix} \gamma_{0,0} &\ldots
    &\gamma_{0,n} \\ \vdots &\ddots &\vdots \\ \gamma_{n,0}
    &\ldots &\gamma_{n,n} \end{matrix} \right\vert \;.
\end{displaymath}

These formulas are easily verified. Indeed, pairing $e_m$ with $w_n$
for $m > n$ one gets
\begin{displaymath}
    \gamma( e_m \, , w_n ) = D_{m-1}^{-1}\, \left\vert \begin{matrix}
    \gamma_{0,0} &\ldots &\gamma_{0,m-1} &\gamma_{0,n} \\ \gamma_{1,0}
    &\ldots &\gamma_{1,m-1} &\gamma_{1,n} \\ \vdots &\ddots &\vdots &\vdots
    \\ \gamma_{m,0} &\ldots &\gamma_{m,m-1} &\gamma_{m,n} \end{matrix}
    \right\vert = 0 \;,
\end{displaymath}
which vanishes because the last column coincides with one of the
other columns. Since $f_n$ is a linear combination of the vectors
$w_{n^\prime}$ with $n^\prime \le n\,$, it follows that $\gamma(e_m
\, , f_n) = 0$ for $m > n$.  The same conclusion for $m < n$ follows
by reversing the roles of $V$ and $W$. Notice that $e_n = 1 \cdot v_n
+ \ldots$ and $f_n = 1 \cdot w_n + \ldots$ by insertion of the factor
$D_{n-1}^{-1}\,$. The non-vanishing pairing matrix elements for $n
\ge 1$ are $\gamma_{n,n} = \gamma(e_n \, , f_n) = \gamma(e_n \, ,
w_n) = D_n / D_{n-1}\,$.

To apply these general formulas to the case under consideration, we
choose the vectors $v_m$ and $w_n$ to be the functions $\omega
\mapsto \omega^m$ resp.\ $\omega \mapsto \omega^{2n}$, and take the
pairing to be given by integration with our measure
$\mathrm{e}^{-\omega} \omega^l\, d\omega:$
\begin{equation}
    \gamma_{m,n} = \int_0^\infty \omega^{m+2n} \mathrm{e}^{-\omega}
    \omega^l\, d\omega = \Gamma(m+2n+l+1) \;.
\end{equation}
Making the identification $e_n \equiv P_n(\omega)$, the general
formula for $e_n$ in (\ref{eq:Gram-Schmidt}) then gives $P_0(\omega)
= 1$ and
\begin{equation}\label{eq:5.6}
    P_n(\omega) = D_{n-1}^{-1} \left\vert \begin{matrix} \Gamma(l+1)
    &\ldots &\Gamma(l+2n-1) &\omega^0\\ \Gamma(l+2) &\ldots &\Gamma(l
    +2n) &\omega^1\\ \vdots &\ddots &\vdots &\vdots \\ \Gamma(l+n+1)
    &\ldots &\Gamma(l+3n-1) &\omega^n \end{matrix} \right\vert
    \qquad (n \ge 1) \;.
\end{equation}
Similarly, identifying $f_n \equiv Q_n(\omega^2)$ we obtain
$Q_0(\omega^2) = 1$ and
\begin{equation}\label{eq:5.7}
    Q_n(\omega^2) = D_{n-1}^{-1} \left\vert \begin{matrix} \Gamma(l+1)
    &\Gamma(l+3) &\ldots &\Gamma(l+2n+1)\\ \vdots &\vdots &\ddots
    &\vdots \\ \Gamma(l+n) &\Gamma(l+n+2) &\ldots &\Gamma(l+3n)\\
    \omega^0 &\omega^2 &\ldots &\omega^{2n} \end{matrix} \right\vert
    \qquad (n \ge 1)\;.
\end{equation}
Using the relation $\Gamma(z+1) = z \, \Gamma(z)$ an easy Gauss
elimination process gives the normalization constant as
\begin{equation}
    D_n = \left\vert \begin{matrix} \Gamma(l+1) &\ldots
    &\Gamma(l+2n+1) \\ \vdots &\ddots &\vdots \\ \Gamma(l+n+1)
    &\ldots &\Gamma(l+3n+1) \end{matrix} \right\vert =
    \prod_{k=0}^n 2^k k! \, (2k+l)! \;.
\end{equation}
From this, note the diagonal pairing matrix elements $h_0 = l!$ and
\begin{equation}\label{eq:5.9}
    \int_0^\infty P_n(\omega)\, Q_n(\omega^2) \, \mathrm{e}^{-\omega}
    \omega^l\, d\omega = h_n = D_n / D_{n-1} = 2^n n! \, (2n+l)!
    \qquad (n \ge 1)\;.
\end{equation}

\subsection{$n$-level correlation functions}\label{sect:V.B}

The $n$-level correlation function $R_n(\omega_1,\ldots,\omega_n)$ in
the present context is defined as
\begin{equation}
    R_n(\omega_1, \omega_2, \ldots, \omega_n) = n! \int\limits_{
    \mathbb{R}_+^N} \sum_{i_1 < i_2 < \ldots < i_n} \delta(\omega_1 -
    \tilde\omega_{i_1})\,\delta(\omega_2 - \tilde\omega_{i_2})\cdots\,
    \delta(\omega_n - \tilde\omega_{i_n})\,d\mu_{N,l}(\tilde\omega_1,
    \ldots, \tilde\omega_N) \;.
\end{equation}
A closed-form expression for it can be given from the bi-orthogonal
polynomials $P_{n^\prime}(\omega)$ and $Q_{n^\prime} (\omega^2)$ for
$0 \le n^\prime \le N$. The result will take its most succinct form
when expressed in terms of the modified functions
\begin{eqnarray}
    \tilde{P}_n(\omega) &:=& (-2)^{-n} n!^{-1}
    \mathrm{e}^{-\omega} P_n(\omega) \;, \label{eq:PtP} \\
    \tilde{Q}_n(\omega) &:=& (-1)^n (2n+l)!^{-1}
    \omega^l\, Q_n(\omega^2) \label{eq:QtQ} \;,
\end{eqnarray}
(the motivation for the sign $(-1)^n$ will become clear later), which
from (\ref{eq:ortho}) and (\ref{eq:5.9}) obey the orthogonality
relations
\begin{equation}\label{eq:new-ortho}
   \int_0^\infty \tilde{P}_m(\omega)\,\tilde{Q}_n(\omega)\,d\omega
   = \delta_{m,n}\;.
\end{equation}
The probability measure (\ref{eq:new2-meas}) expressed by these
functions takes the form
\begin{displaymath}
    d\mu_{N,l}(\omega_1,\ldots,\omega_N) = \frac{1}{N!}\  \mathrm{Det}
    \,\big( \tilde{P}_{i-1}(\omega_j) \big) \, \mathrm{Det}\, \big(
    \tilde{Q}_{i-1}(\omega_j)\big) \prod\nolimits_k d\omega_k \;.
\end{displaymath}

Now, by using the multiplicative property of the determinant, we can
also write
\begin{equation}\label{eq:kernel-meas}
    d\mu_{N,l}(\omega_1,\ldots,\omega_N) = \frac{1}{N!} \, \mathrm{Det}
    \,\big(K_N(\omega_i \, , \omega_j) \big)_{i,j=1,\ldots,N}
    \prod\nolimits_k d\omega_k \;,
\end{equation}
where the kernel $K(\omega_i\, , \omega_j)$ is defined by
\begin{equation}\label{eq:kernel}
    K_N(\omega_i\, , \omega_j) = \sum_{n=0}^{N-1}\tilde{P}_n(\omega_i)
    \,\tilde{Q}_n(\omega_j)\;.
\end{equation}
From the orthogonality relations (\ref{eq:new-ortho}) this kernel has
the reproducing property
\begin{equation}\label{eq:reproduce}
    \int_0^\infty K_N(\omega_i\, ,\omega)\, K_N(\omega,\omega_j)\,
    d\omega = K_N(\omega_i\, ,\omega_j) \;,
\end{equation}
and the trace
\begin{equation}\label{eq:trace}
    \int_0^\infty K_N(\omega,\omega)\, d\omega = N \;.
\end{equation}

To proceed further, take notice of the relation
\begin{displaymath}
    \int_0^\infty \left\vert \begin{matrix} K_N(\omega_1,\omega_1)
    &\ldots &K_N(\omega_1,\omega_n)\\ \vdots &\ddots &\vdots\\
    K_N(\omega_n,\omega_1) &\ldots &K_N(\omega_n,\omega_n) \end{matrix}
    \right\vert d\omega_n = (N-n+1) \, \left\vert \begin{matrix} K_N(\omega_1,
    \omega_1) &\ldots &K_N(\omega_1,\omega_{n-1})\\ \vdots &\ddots &\vdots
    \\ K_N(\omega_{n-1},\omega_1)&\ldots &K_N(\omega_{n-1},\omega_{n-1})
    \end{matrix} \right\vert \;,
\end{displaymath}
which is proved by expanding the determinant with respect to the last
row or column and exploiting the properties (\ref{eq:reproduce}) and
(\ref{eq:trace}). Using it, an inductive procedure starting from
$R_N(\omega_1 ,\ldots,\omega_N) = \mathrm{Det}\, \big( K_N(\omega_i
\,,\omega_j) \big)_{i,j=1,\ldots, N}$ gives the $n$-level correlation
functions as
\begin{equation}\label{eq:correls}
    R_n(\omega_1,\ldots,\omega_n) = \mathrm{Det}\,\big( K_N(\omega_i\,
    ,\omega_j) \big)_{i,j = 1, \ldots, n} \;.
\end{equation}
Thus the correlations are those of a determinantal process and are
completely determined by the kernel $K_N(\omega_i\, ,\omega_j)$. The
remaining discussion therefore focuses on this kernel, but first we
make another preparatory step.

\subsection{Contour integral representation}\label{sect:V.C}

We are now going to show that the functions $\tilde{P}_n(\omega)$ and
$\tilde{Q}_n(\omega)$ have expressions as complex contour integrals:
\begin{eqnarray}
     \tilde{P}_n(\omega) &=& \oint_{S_\epsilon (1)}
     \mathrm{e}^{-\omega u} \, (1 - u^{-2})^{-n-1} u^{l-2} du/
     \pi\mathrm{i} \label{eq:P-def} \;, \\ \tilde{Q}_n(\omega) &=&
     \oint_{S_{\epsilon}(0)} \mathrm{e}^{\omega v} \,
     (1-v^{-2})^n v^{-l-1} dv / 2\pi\mathrm{i} \label{eq:Q-def} \;.
\end{eqnarray}
Both integrals are over circles in the complex plane with radius
$\epsilon$ and counterclockwise orientation; the first circle is
centered at $u = 1$ and has radius $\epsilon < 2$ (to avoid the
singularity at $u = -1$), the second one is centered at $v = 0\,$.

Our proof of these expressions for $\tilde{P}_n(\omega)$ and
$\tilde{Q}_n(\omega)$ will be indirect and in two steps. First, we
establish some information on power series. In the case of
$\tilde{Q}_n(\omega)$ we insert the power series of the exponential
function $\mathrm{e}^{\omega v}$, use the binomial expansion of $(1 -
v^{-2})^n$, and compute a residue to obtain
\begin{equation}\label{eq:Q-pol}
    \tilde{Q}_n(\omega) = \sum_{k=0}^n \binom{n}{k}
    \frac{(-1)^k \omega^{2k+l}}{(2k+l)!} \;.
\end{equation}
In the case of $\tilde{P}_n(\omega)$, calculating the residue at $u =
1$ we have that
\begin{equation}\label{eq:P-pol}
     \tilde{P}_n(\omega) = \frac{2}{n!} \frac{d^n}{du^n}\left(
     \frac{\mathrm{e}^{-\omega u} u^{n+l-1}}{(1+u^{-1})^{n+1}}\right)
     \Bigg\vert_{u = 1} \;.
\end{equation}
In both cases, defining $P_n(\omega)$ and $Q_n(\omega^2)$ by the
reverse of the relations (\ref{eq:PtP}, \ref{eq:QtQ}), we see from
(\ref{eq:Q-pol}, \ref{eq:P-pol}) that these are polynomials of degree
$n$ in $\omega$ resp.\ $\omega^2$ and that the highest-degree term
($\omega^n$ resp.\ $\omega^{2n}$) has coefficient one.

Recall now from Sect.\ \ref{sect:V.A} that, given these properties,
the polynomials $P_n(\omega)$ and $Q_n(\omega^2)$ are completely
determined by the orthogonality relations (\ref{eq:ortho}) for $m
\not= n$. Via (\ref{eq:PtP},\ref{eq:QtQ}) the latter are in
one-to-one correspondence with the orthogonality relations
(\ref{eq:new-ortho}) (still for $m \not= n$). Therefore, defining
\begin{equation}\label{eq:5.23}
    \tilde{I}_{m,n} = \int_0^\infty \tilde{P}_m(\omega)\,
    \tilde{Q}_n(\omega) \, d\omega \;,
\end{equation}
the second and final step of our proof is to show that $\tilde{I}_{m
,n} = 0$ for $m \not= n$.

To that end, we insert the expressions (\ref{eq:P-def},
\ref{eq:Q-def}) into (\ref{eq:5.23}). The $\omega$-dependence then is
$\mathrm{e}^{- \omega (u-v)}$ with $u\in S_\epsilon(1)$ and $v \in
S_\epsilon(0)$. Taking the radius $\epsilon$ to be very small
$(\epsilon \ll 1)$, we have that $\mathrm{e}^{-\omega (u-v)}$
decreases rapidly as $\omega$ goes to $+\infty$. Therefore, the
integral over $\omega$ exists, and we may interchange the order of
integrations. Doing first the $\omega$-integral,
\begin{displaymath}
    \int_0^\infty \mathrm{e}^{-\omega(u-v)}\,d\omega =
    \frac{1}{u-v}\;,
\end{displaymath}
the remaining contour integrals for $\tilde{I}_{m,n}$ defined by
(\ref{eq:5.23}) are
\begin{displaymath}
    \tilde{I}_{m,n} = \oint_{S_\epsilon(1)}\frac{u^{l-2}}{(1 -
    u^{-2})^{m+1}}\left(\oint_{S_{\epsilon}(0)}\frac{(1-v^{-2})^n\,
    dv}{v^{l+1} (v-u)}\right)\frac{du}{2\pi^2} \;.
\end{displaymath}
To simplify the inner integral over $v$ we use the identity
\begin{displaymath}
    \left( \frac{1-v^{-2}}{1-u^{-2}} \right)^n = 1 -
    \frac{v^2-u^2}{v^2(1-u^2)} \sum_{k=0}^{n-1} \left(
    \frac{1-v^{-2}}{1-u^{-2}} \right)^k \;.
\end{displaymath}
Inserting this into the expression for $\tilde{I}_{m,n}$ we see that
the terms in the $k$-sum do not contribute as the residue at $v = 0$
vanishes for all of those terms. Doing the $v$-integral for the first
term on the right-hand side, we get
\begin{displaymath}
    \oint_{S_\epsilon (0)} v^{-l-1} (v - u)^{-1} dv = - 2\pi
    \mathrm{i}\, u^{-l-1} \;,
\end{displaymath}
so the remaining $u$-integral is
\begin{displaymath}
    \tilde{I}_{m,n} = (\pi\mathrm{i})^{-1} \oint_{S_\epsilon (1)}
    (1 - u^{-2})^{n-m-1} u^{-3} du \;.
\end{displaymath}
This integrand is holomorphic near $u = 1$ for $m < n\,$, and the
integral therefore vanishes in that case. For $m > n$ we use the
invariance of the integration form under $u \to -u$ to write
$\tilde{I}_{m,n}$ as an integral over a sum of two circles:
\begin{displaymath}
    \tilde{I}_{m,n} = \frac{1}{2\pi\mathrm{i}}
    \oint_\gamma\frac{u^{2m-2n-1} du}{(u^2 - 1)^{m-n+1}}\;,
    \qquad \gamma = S_\epsilon (1) + S_\epsilon(-1) \;.
\end{displaymath}
The integrand in this case is holomorphic near $u = 0\,$.  In the
punctured plane $\mathbb{C} \setminus \{1 , -1 \}$ the chain
$S_\epsilon(1) + S_\epsilon(-1)$ is homologous to the circle at
infinity, where the integrand vanishes. Therefore the integral again
is zero. This proves that $\tilde{I}_{m,n} = 0$ for $m \not= n$,
which in turn completes our proof that the contour integrals
(\ref{eq:P-def}) and (\ref{eq:Q-def}) are the same as the functions
$\tilde{P}_n(\omega)$, $\tilde{Q}_n(\omega)$ defined from
(\ref{eq:5.6}, \ref{eq:5.7}) by (\ref{eq:PtP}, \ref{eq:QtQ}). As a
final check, note that
\begin{displaymath}
    \tilde{I}_{n,n} = (\pi\mathrm{i})^{-1}
    \oint_{S_\epsilon (1)} \frac{du}{u(u^2-1)} = 1 \;,
\end{displaymath}
which is what it ought to be in view of (\ref{eq:new-ortho}).

Now we harvest a major benefit from the contour integral
representations (\ref{eq:P-def}) and (\ref{eq:Q-def}): using these,
we can carry out the sum in the definition (\ref{eq:kernel}) of the
kernel $K_N$ as a geometric sum.  The result is a double contour
integral:
\begin{eqnarray}
    K_N(\omega_1,\omega_2) &=& \oint\limits_{S_\epsilon(1)} du
    \oint\limits_{S_\epsilon(0)}dv\,\, F_N(u,v\,;\omega_1,\omega_2)
    \;, \label{contourK}\\ F_N(u,v\,;\omega_1,\omega_2) &=&
    \frac{1}{2\pi^2}\, \mathrm{e}^{-\omega_1 u +\omega_2 v}\,
    \frac{u^l \, v^{-l+1}}{u^2-v^2}\left(\left(\frac{1-v^{-2}}
    {1-u^{-2}}\right)^N - 1\right) \label{eq:FN} \;.
\end{eqnarray}
This exact expression represents the complete solution of our
problem. We will now use it to determine the large-$N$ asymptotics in
the bulk and at the hard edge $\omega = 0\,$.

\subsection{Asymptotics in the bulk}\label{sect:V.D}

The kernel on the diagonal $\omega_1 = \omega_2$ is the same as the
1-level function, $R_1(\omega) = K_N(\omega,\omega)$; see
(\ref{eq:correls}). We already know from Sect.\ \ref{sect:cubic} the
asymptotics of $R_1(\omega) \equiv \rho(\omega)$ in the bulk:
introducing the scaling variable $x = \omega / N$ (formerly $x =
\omega\tau / N$), this is
\begin{displaymath}
    \lim_{N \to \infty} K_N(Nx,Nx) = \rho_\infty(x) \;,
\end{displaymath}
with $\rho_\infty(x)$ given by (\ref{eq:soln-rho}).  In the present
subsection we are going to demonstrate that the scaling limit of the
kernel $K_N(\omega_1, \omega_2)$ off the diagonal leads to
sine-kernel universality for all level correlation functions:
\begin{equation}\label{eq:sine-kernel}
    \lim_{N \to \infty} R_n(Nx + \omega_1, \ldots, Nx + \omega_n) =
    \mathrm{Det}\left(\frac{\sin\big(\pi \rho_\infty(x) (\omega_i-
    \omega_j)\big)}{\pi(\omega_i-\omega_j)}\right)_{i,j=1,\ldots,n}\;,
\end{equation}
as is expected for systems in the universality class of the Gaussian
Unitary Ensemble. As a corollary, we will obtain an independent
confirmation of the result (\ref{eq:soln-rho}).

Looking at the integral representation (\ref{contourK}) one might
think that the large-$N$ limit could be taken by applying the
saddle-point method to that integral.  However, as we shall see, the
dominant saddle points lie on the line $u = v$ where the integrand
has a singularity of type $0/0$ which, albeit removable, complicates
the saddle-point evaluation.

Therefore, rather than calculating $K_N(\omega_1,\omega_2)$ directly,
we look at the product $(\omega_1 - \omega_2)\, K_N(\omega_1,
\omega_2)$. Using the relation $(\omega_2 - \omega_1) \, \mathrm{e}^{
\omega_2 v - \omega_1 u} = (\partial_v + \partial_u)\, \mathrm{e}^{
\omega_2 v - \omega_1 u}$ and partially integrating, we rewrite
(\ref{contourK}) as
\begin{eqnarray}
    (\omega_1-\omega_2)\, K_N(\omega_1,\omega_2) &=&
    \oint\limits_{S_\epsilon(1)} du \oint\limits_{S_\epsilon(0)}dv\,
    \,\tilde{F}_N(u,v\,;\omega_1,\omega_2) \label{eq:new-K} \;, \\
    \tilde{F}_N(u,v\,;\omega_1,\omega_2)&=& \frac{1}{2\pi^2}\,
    \mathrm{e}^{-\omega_1 u +\omega_2 v}\left(\frac{\partial}{\partial u}
    + \frac{\partial}{\partial v}\right) \frac{u^l \, v^{-l+1}}{u^2-v^2}
    \left(\left(\frac{1-v^{-2}}{1-u^{-2}}\right)^N - 1\right)\;,
\end{eqnarray}
which constitutes the starting point for the following analysis.

In preparation for taking the limit $N \to \infty\,$, we set
$\omega_1 = Nx + \omega$ and $\omega_2 = Nx + \tilde\omega\,$. The
deciding factor in the integrand of $(\omega_1 - \omega_2)\,
K_N(\omega_1\, , \omega_2)$ in the large-$N$ limit will then be
\begin{displaymath}
    \exp\left( - Nx(u-v) + N \log(1 - v^{-2}) - N \log(1 - u^{-2})
    \right) \;,
\end{displaymath}
leading to the saddle-point equation
\begin{equation}\label{eq:SPEQ}
    \varphi(u) = x = \varphi(v) \;, \qquad \varphi(w) = - \partial_w
    \log (1 - w^{-2}) = \frac{-2}{w (w^2-1)} \;.
\end{equation}
Notice that $\varphi$ is related to our function (\ref{eq:cubic}) by
$f(w - 1) = \varphi(w)$. A comprehensive study of the equation $f(w)
= z$ and its solutions for $w$ was made in Sect.\ \ref{sect:cubic}.
From there we know that the saddle-point equation $\varphi(w) = x$
has three solutions in general, and for $0 < x \le b = 3 \sqrt{3}$
these are
\begin{equation}\label{eq:5.29}
    w_\alpha(x) = - x^{-1/3} \mathrm{e}^{- 2\pi\mathrm{i} \alpha / 3}
    (1 + \sqrt{1 - x^2/b^2})^{1/3} - x^{-1/3} \mathrm{e}^{2\pi
    \mathrm{i} \alpha /3} (1 - \sqrt{1 - x^2/b^2})^{1/3} \qquad
    (\alpha = 1,-1,0) \;.
\end{equation}
In the range of interest ($0 < x < b$) the first two solutions,
$w_{\pm 1}(x)$, are complex conjugates of each other while the third
one, $w_0(x)$, is negative.  Expanding the logarithm of $(1-v^{-2})^N
/ (1-u^{-2})^N$ to second order around a pair of saddle points
$w_\alpha\, , w_\beta$ we encounter the Gaussian
\begin{displaymath}
    \exp\left({\textstyle{\frac{1}{2}}} N \varphi^\prime(w_\alpha(x))
    (\delta u)^2 - {\textstyle{\frac{1}{2}}} N \varphi^\prime(w_\beta(x))
    (\delta v)^2 \right) \;, \qquad \varphi^\prime(w) = \frac{6w^2 -
    2}{w^2 (w^2-1)^2} \;.
\end{displaymath}
For the negative saddle point one has $\varphi^\prime(w_0(x)) > 0\,$,
so its path of steepest descent would be perpendicular to the real
axis in the case of $u$ and along the real axis in the case of $v$.
The latter is inconsistent with the original integration contour for
$v$ being $S_\epsilon(0)$. In the former case, $w_0(x) < -1 $ is
inaccessible because of the singularity of $(1 - u^{-2})^{-N}$
intervening at $u = -1$. Thus this saddle point is irrelevant for
present purposes and may be discarded.

We now make another preparation of the saddle-point evaluation of the
integral, by investigating the behavior of the integrand near the two
remaining saddle points. We set
\begin{displaymath}
    u = w_\alpha(x) + N^{-1/2} \delta u\;, \quad v = w_\beta(x)
    + N^{-1/2} \delta v \qquad (\alpha, \beta = \pm 1) \;,
\end{displaymath}
and first look at the diagonal case, $\alpha = \beta\,$. Using the
identity
\begin{equation}\label{eq:auxiliary}
    \frac{1}{u-v} \left( \frac{\partial}{\partial u} + \frac{\partial}
    {\partial v} \right) \left( \frac{1-v^{-2}}{1- u^{-2}} \right)^N =
    2N\, \frac{u^2 + uv + v^2 - 1}{u(u^2-1)\,v(v^2-1)} \left(\frac{1 -
    v^{-2}}{1 - u^{-2}} \right)^N \;,
\end{equation}
we find the scaling limit of the integrand $\tilde{F}_N$ to be
\begin{eqnarray*}
    &&\lim_{N \to \infty} N^{-1} \tilde{F}_N(w_\alpha(x) + N^{-1/2}
    \delta u\, , w_\alpha(x) + N^{-1/2} \delta v \, ; Nx + \omega
    \, , Nx + \tilde\omega)\\ &&\hspace{2cm} = (2\pi)^{-2} \mathrm{e}^{
    -w_\alpha(x)(\omega-\tilde\omega)}\varphi^\prime(w_\alpha(x))\,
    \mathrm{e}^{\frac{1}{2} \varphi^\prime(w_\alpha(x))
    (\delta u^2 - \delta v^2)} \;.
\end{eqnarray*}
The same limit in the off-diagonal case $(\alpha \not= \beta)$
vanishes. Indeed,
\begin{displaymath}
    w_\alpha^2 + w_\alpha w_\beta + w_\beta^2 - 1 =
    \frac{w_\alpha(w_\alpha^2 - 1) - w_\beta (w_\beta^2 -1)}
    {w_\alpha - w_\beta} = \frac{-2}{w_\alpha - w_\beta} (x^{-1} -
    x^{-1}) = 0 \quad (\alpha \not= \beta)\;,
\end{displaymath}
and therefore the factor in the numerator on the right-hand side of
(\ref{eq:auxiliary}) gives zero.

\begin{figure}
    \input{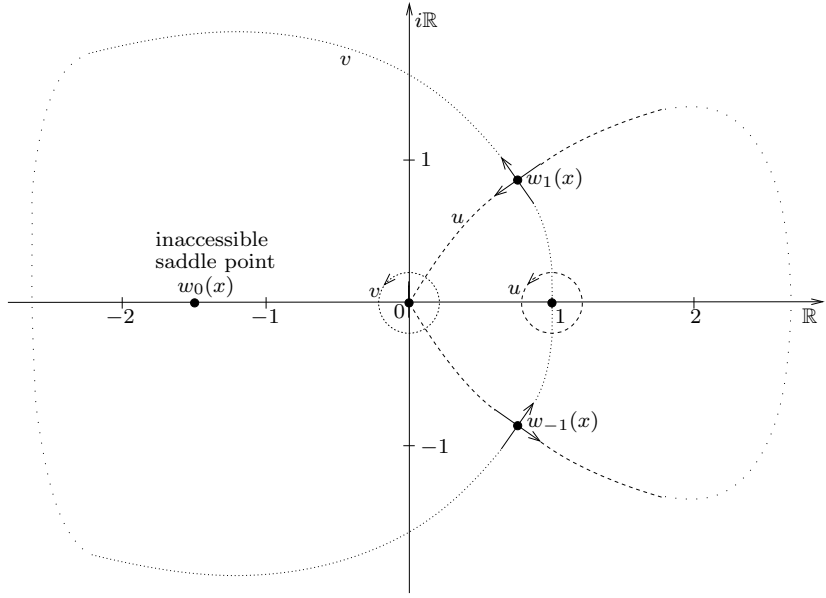}
    \caption{Sketch of the saddle points for the case of $x = 1\,$. By
    deforming the original contours, which are small circles around the
    singular points $u = 1$ and $v = 0\,$, one arranges for the contours
    of integration to pass through the saddle points $w_{\pm 1}(x)$ in
    the direction of steepest descent. Away from the saddle points
    the deformed contours are drawn as paths of constant phase, which
    interpolate between different zeroes of the integrand: they run
    between $0$ and $+\infty$ for $u\,$, and between $1$ and $-\infty$
    for $v\,$.} \label{fig:3}
\end{figure}

We now deform the contours of integration as indicated in Fig.\
\ref{fig:3}. The deformed contours pass through the saddle points
$w_{\pm 1}$ but miss the saddle point $w_0\,$. At $w_{\pm 1}$ the
paths of steepest descent for $u$ and $v$ cross at right angles,
valleys in one case being mountains in the other case and vice versa.

Next we do the Gaussian integrals. Given the counterclockwise
orientations of the original contours $S_\epsilon(1)$ resp.\
$S_\epsilon(0)$, and taking into account the directions of the paths
of steepest descent, we get
\begin{eqnarray*}
    \int \mathrm{e}^{\frac{1}{2} \varphi^\prime(w_{+1})\, \delta u^2}
    d(\delta u) &=& \sqrt{2\pi}\, |\varphi^\prime(w_{+1})|^{-1/2}
    \mathrm{e}^{\frac{1}{2}(-\pi\mathrm{i} - \mathrm{i}
    \arg \varphi^\prime(w_{+1}))} \;,\\\int \mathrm{e}^{-\frac{1}{2}
    \varphi^\prime(w_{+1})\, \delta v^2} d(\delta v) &=&
    \sqrt{2\pi}\, |\varphi^\prime(w_{+1})|^{-1/2}\mathrm{e}^{\frac{1}{2}
    (2\pi\mathrm{i} - \mathrm{i}\arg \varphi^\prime(w_{+1}))} \;.
\end{eqnarray*}
The product of these two integrals is $2\pi\mathrm{i}/ \varphi^\prime
(w_{+1})$. The same calculation for the other saddle $w_{-1}$ gives
$-2\pi \mathrm{i} / \varphi^\prime(w_{-1})$. Thus, putting the
factors together and summing over the contributions from diagonal
pairs of saddle points $(\alpha = \beta)$ we obtain
\begin{displaymath}
    (\omega - \tilde\omega)\, \lim_{N\to\infty} K_N(Nx + \omega ,
    Nx + \tilde\omega) = \frac{1}{2\pi\mathrm{i}} \big( \mathrm{e}^{
    -w_{-1}(x)(\omega - \tilde\omega)} -\mathrm{e}^{-w_{+1}(x)
    (\omega - \tilde\omega)}\big) \;.
\end{displaymath}
Since $\mathfrak{Re}\, w_{+1} = \mathfrak{Re}\, w_{-1}$ and
$\mathfrak{Im}\, w_{+1} = - \mathfrak{Im}\, w_{-1}$ this means that,
with $\mathfrak{Im}\, w_{+1}(x) =: \pi \rho_\infty(x)$, we have
\begin{displaymath}
    \lim_{N\to\infty} K_N( Nx + \omega , Nx + \tilde\omega) =
    \mathrm{e}^{-\mathfrak{Re}\, w_{\pm 1}(x) (\omega - \tilde\omega)}
    \times\frac{\sin\big(\pi \rho_\infty(x) (\omega - \tilde\omega)\big)}
    {\pi(\omega -  \tilde\omega)} \;.
\end{displaymath}
The exponential factor $\mathrm{e}^{-\mathfrak{Re}\, w_{\pm 1}(x)
(\omega - \tilde\omega)}$ drops out when forming the determinant on
the right-hand side of (\ref{eq:correls}). Thus we arrive at the
universal sine-kernel (or GUE) correlation functions
(\ref{eq:sine-kernel}).

Setting $\omega = \tilde\omega$ notice the special result
$\lim_{N\to\infty} K_N(Nx , Nx) = \rho_\infty(x)\,$. Since the kernel
on the diagonal is none other than the 1-level function,
$K_N(\omega,\omega) = \rho(\omega)$, this gives another determination
of the large-$N$ level density $\rho_\infty\,$. From (\ref{eq:5.29})
one sees that $\rho_\infty(x) = \pi^{-1} \mathfrak{Im}\, w_{+1}(x)$
agrees with our earlier result (\ref{eq:soln-rho}).

\subsection{Asymptotics near $\omega = 0$}\label{sect:V.E}

At the lower edge ($\omega = 0$) of the spectrum, a new type of
behavior is expected to emerge. This behavior, as we shall see
presently, occurs on a scale $\omega \sim N^{-1/2}$.

To exhibit the scaling limit near $\omega = 0\,$, it is best to send
the integration variables $u ,v$ to their reciprocals, $u \to u^{-1}$
and $v \to v^{-1}$. Then $du \to - u^{-2} du\,$, $dv \to - v^{-2}
dv\,$, and the integration contour for $v$ has its radius inverted
and orientation reversed, $S_\epsilon(0) \to - S_{1/\epsilon}(0)$.
However, since the integrand is holomorphic in $v$ on $\mathbb{C}
\setminus \{ 0 \}$ we may return to the original radius $\epsilon$
(or any other radius, for that matter). In the case of $u$ we take
the radius $\epsilon$ of $S_\epsilon(1)$ to be very small.  Then
inversion $u \to u^{-1}$ sends $S_\epsilon(1)$ to itself (or, in any
case, to the same homology class on $\mathbb{C} \setminus \{ 1 \})$,
with no change of orientation. Altogether, then, carrying out the
transformation $(u,v) \to (u^{-1},v^{-1})$ the integral
representation (\ref{contourK}) continues to hold true if we make the
replacement
\begin{displaymath}
    F_N(u,v\,;\omega_1,\omega_2)\to - u^{-2} v^{-2} F_N(u^{-1},v^{-1};
    \omega_1 , \omega_2) =\frac{1}{2\pi^2}\, \mathrm{e}^{-\omega_1 / u
    + \omega_2 / v}\, \frac{u^{-l}\, v^{l-1}}{u^2-v^2}\left(\left(
    \frac{1-v^2} {1-u^2}\right)^N - 1\right) \;.
\end{displaymath}

Next, as another preparation for taking the limit $N \to \infty$, we
deform the $u$-contour $S_\epsilon(1)$ to some axis parallel to the
imaginary axis. The deformed contour crosses the real axis between $u
= 0$ and $u = 1$ and is directed from $u = +\mathrm{i}\infty$ to $u=
- \mathrm{i}\infty$. We also reverse the direction of integration for
$u$ and change the overall sign of the integral.

\begin{figure}
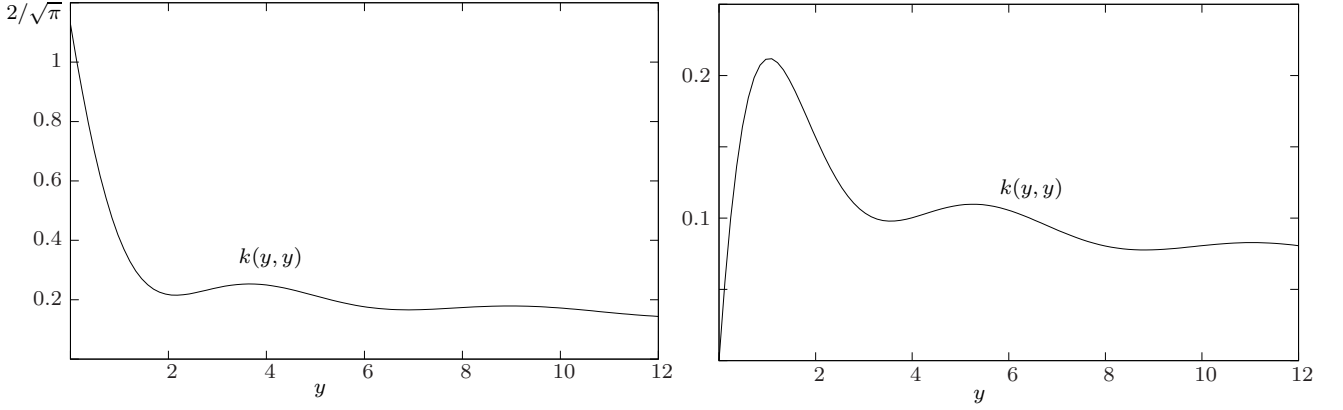

    \parbox[r]{8.7cm}{\input{fig4.pstex_t}}
    \parbox[r]{8.7cm}{\input{fig2.pstex_t}}
    \caption{The graph of the scaling function $k(y,y)$ for the case of
    $l = 0$ (left) and $l = 1$ (right).}
    \label{fig:2}
\end{figure}

Then we set $\omega_j = N^{-1/2} y_j$ and rescale $u \to N^{-1/2} u$
and $v \to N^{-1/2} v$ accordingly. Again, in view of the analytic
properties of the integrand we can keep the integration contours
fixed while rescaling. Because the $u$-integral converges at infinity
we have a good limit
\begin{displaymath}
    \lim_{N \to \infty} (1 - u^2 / N)^{-N} = \exp(u^2) \;.
\end{displaymath}
In total, we thus obtain the following scaling limit for our kernel
$K_N:$
\begin{equation}\label{eq:asymptK}
    k(y_1,y_2) := \lim_{N\to\infty} N^{-1/2} K_N(N^{-1/2} y_1 ,
    N^{-1/2} y_2) = \frac{1}{2\pi^2} \int\limits_{\mathrm{i}\mathbb{R}
    + \epsilon} du \oint\limits_{\mathrm{U}_1} \frac{dv}{v} \,\,
    \mathrm{e}^{- y_1 / u + y_2 / v}\, (v/u)^{l} \,
    \frac{\mathrm{e}^{u^2 - v^2} - 1}{u^2 - v^2} \;,
\end{equation}
where $\mathrm{U}_1 \equiv S_1(0)$ means the unitary numbers, and
$\mathrm{i}\mathbb{R} + \epsilon$ is the imaginary axis translated by
$\epsilon > 0$ into the right half of the complex plane.  Plots of
the scaling function $k(y,y)$ for $l = 0, 1$ are shown in Fig.\
\ref{fig:2}. Using the method of saddle-point evaluation as in Sect.\
\ref{sect:V.C} one can show that this function behaves as $k(y,y)
\sim y^{-1/3}$ for large $y$.

Taking the same scaling limit for the functions $\tilde{P}_N(\omega)$
and $\tilde{Q}_N(\omega)$ in (\ref{eq:P-def}) and (\ref{eq:Q-def})
one gets
\begin{eqnarray}
    p(y) &=& \lim_{N \to \infty} N^{-(l-1)/2} \tilde{P}_N(N^{-1/2} y) =
    \frac{1}{\pi\mathrm{i}} \int\limits_{\mathrm{i}\mathbb{R}+\epsilon}
    \mathrm{e}^{u^2 - y/u}\, u^{-l} du \;, \label{eq:p(y)}\\
    q(y) &=& \lim_{N \to \infty} N^{l/2} \tilde{Q}_N(N^{-1/2} y)
    = \frac{1}{2\pi\mathrm{i}} \oint\limits_{\mathrm{U}_1}
    \mathrm{e}^{-v^2 + y/v}\, v^{l-1} dv \label{eq:q(y)} \;.
\end{eqnarray}
Both functions have convergent series expansions:
\begin{equation}\label{eq:5.28}
    p(y) = \sum_{n=0}^{\infty} \frac{(-y)^n}{n!\,\Gamma((l+n+1)/2)}
    \;, \qquad q(y) = y^l \sum_{n=0}^{\infty}\frac{(-y^2)^n}
    {n!\, (2n+l)!} \;.
\end{equation}
The expansion for $q(y)$ can be obtained either directly from
(\ref{eq:q(y)}), or by taking the limit $N \to \infty$ in
(\ref{eq:Q-pol}).  In the case of $p(y)$, the earlier formula
(\ref{eq:P-pol}) is not suitable; rather, in order to verify
(\ref{eq:5.28}) for $p(y)$ one expands the integrand of
(\ref{eq:p(y)}) in powers of $y\,$, makes use of $\mathfrak{Re}\, u =
\epsilon > 0$ to write
\begin{displaymath}
    u^{-n-l} = (n+l-1)!^{-1} \int_0^\infty \mathrm{e}^{-tu} \,
    t^{n+l-1} \, dt \qquad (n+l > 0) \;,
\end{displaymath}
does the Gaussian $u$-integral by completing the square, and uses the
duplication formula for the Gamma function.

\bigskip\noindent{\bf Acknowledgement}. We thank our colleague
T.\ Kriecherbauer for helpful discussions. This work has been
supported by funding from the Deutsche Forschungsgemeinschaft (TRR
12).

\bigskip\noindent{\bf Note added.} After submission of this
manuscript, P.\ Forrester pointed out to us that the joint eigenvalue
distribution derived and analyzed here falls in a broad class of
models solved by Borodin \cite{borodin}. Borodin's expression for the
kernel $K_N(x,y)$ is equivalent to ours by old work of Konhauser
\cite{konhauser}.  The mathematical results of Konhauser were first
introduced into random matrix physics by Muttalib \cite{muttalib}, who
suggested to use them for an approximate treatment of the statistics
of transmission eigenvalues of disordered conductors.

\end{document}